# Seismic P- and S-wave attenuation inversion in time-frequency domain based on Kramers-Kronig relations

Shengyi Wang[1,2], Xuehua Chen[1,2], Xingyu Luo[2], Kaixing Huang[2]

[1]Chengdu University of Technology, State Key Laboratory of Oil & Gas Reservoir Geology and Exploitation, Chengdu 610059, China.

[2]Chengdu University of Technology, Key Laboratory of Earth Exploration & Information Techniques of Ministry of Education, Chengdu 610059, China.

E-mail: wangsy_min@163.com; chen_xuehua@163.com; luoxingyu_sco1031@163.com; 13330645617@163.com.

# Abstract

Seismic attenuation (inverse quality factor) inversion is important for gas reservoir prediction. Two key issues in seismic attenuation inversion are the development of a physically consistent reflection coefficient equation and stable inversion of seismic attenuation from seismic data. To address these issues, this study, within the framework of isotropic linear viscoelastic media, starts from the Kramers-Kronig relations and expresses the viscoelastic stiffness matrix as a function of seismic attenuation. Under the assumptions of weak attenuation and small elastic and attenuation contrasts across the interface, a frequency-dependent PP-wave reflection coefficient equation explicitly containing P- and S-wave attenuation terms is derived using scattering theory. The derived equation not only satisfies the causality constraint but also preserves a compact mathematical form. Based on this equation, P- and S-wave attenuation are estimated by a time-frequency domain Bayesian inversion. To improve resolution and stability, the inversion adopts a recursive frequency strategy with half-power overlapping Gaussian windows, in which the result at the previous frequency serves as the prior model for the next frequency, and prior models are built from well-log rock-physics modeling to provide low-frequency constraints. Synthetic seismic data tests show that the inversion resolution and accuracy improve as the inversion frequency increases and that the method remains robust at a signal-to-noise ratio of 10 dB. Application to field seismic data further demonstrates that the P-wave attenuation anomalies effectively identify gas reservoirs, whereas S-wave attenuation is insensitive to gas reservoirs. This study provides a new approach for extracting P- and S-wave attenuation information from seismic data and achieving high-resolution prediction of gas reservoirs.

# 1 Introduction

A large number of rock physics experiments have demonstrated that fluid-saturated rocks usually exhibit significant seismic dispersion and attenuation. Moreover, the dispersion and attenuation characteristics are strongly related to fluid type, gas saturation, and pore structure (Batzle et al., 2006; Adam et al., 2009; Tisato and Quintal, 2013; Pimienta et al., 2015; Chapman et al., 2016; Mikhaltsevitch et al., 2016; Yin et al., 2017; Zhao et al., 2021). Therefore, seismic dispersion and attenuation are sensitive to reservoir fluids, and the differences in dispersion and attenuation between reservoirs and non-reservoirs can be used for reservoir prediction.

Frequency-dependent AVO (FDAVO) inversion is an important approach for reservoir prediction. From the perspective of reflection coefficient modeling, current FDAVO methods can be broadly divided into two categories. One category starts from the scattering response of layered media, establishes complex reflection coefficients that vary with frequency and angle, and thereby extracts scattering-based FDAVO attributes (Kumar et al., 2020; Zhao et al., 2021). Scattering-based FDAVO treats the reservoir as a layered elastic medium, and its frequency dependence mainly arises from intra-layer propagation time delays, interlayer interference, and P-SV mode conversion (Foster et al., 2017). Kumar et al. (2020) represented the FDAVO response as a complex amplitude varying with angle and frequency, controlled by the velocities and thicknesses of individual layers and the conventional AVO response at all interfaces, and used well-log forward modeling to identify tuning frequencies before performing seismic FDAVO analysis. Zhao et al. (2021) further considered P-SV conversion and proposed a second-order FAVO gradient attribute for thin-bed identification. This type of method provides an important means for layered reservoir and thin-bed identification, and its inversion targets are mainly elastic reflectivities or second-order scattering attributes. The other category introduces frequency into approximate reflection coefficients derived from the Zoeppritz equation to characterize seismic reflection dispersion, and combines time-frequency analysis methods to invert modulus dispersion attributes (Wilson et al., 2009). Previous studies have shown that the above FDAVO dispersion attributes and scattering-based FDAVO attributes can be used for gas reservoir prediction (Wu et al., 2014; Zong et al., 2016; Guo et al., 2022a; Luo et al., 2023; Zhao et al., 2023; Zhao et al., 2026).

However, the Zoeppritz-type frequency-extended method mainly focuses on modulus dispersion attributes and does not explicitly establish the relationship between seismic attenuation and frequency-dependent reflection coefficients. The frequency dependence of scattering-based FDAVO mainly comes from scattering time delays and interlayer interference rather than an explicit attenuation parameter. Therefore, to directly invert seismic attenuation, it is still necessary to construct a frequency-dependent reflection coefficient that explicitly contains attenuation parameters.

In addition to the above two types of FDAVO attributes, the inverse quality factor

($Q^{-1}$), which quantitatively describes seismic attenuation in viscoelastic media, is also an important parameter for reservoir prediction and fluid identification (Korneev et al., 2004; Dvorkin and Mavko, 2006; Innanen, 2011; Mavko et al., 2020; Ahmed et al., 2023; Kuai et al., 2026; Qadrouh et al., 2026). Establishing a reasonable PP-wave reflection coefficient equation in viscoelastic media is the key for seismic attenuation inversion (Borcherdt, 2009; Moradi and Innanen, 2015; Chen, 2020b; Ahmed et al., 2023). Borcherdt (2009) extended Snell's law in elastic media to the complex generalized Snell's law in viscoelastic media and developed a complete theory of reflection and transmission for inhomogeneous plane waves in layered viscoelastic media. Under the weak attenuation assumption ($Q^{-1} \ll 1$), Moradi and Innanen (2015, 2016) further derived scattering potentials for different viscoelastic wave modes by extending Borcherdt's theory to arbitrary 3D heterogeneous media, and proposed a linearized viscoelastic PP-wave AVO approximation incorporating attenuation effects. These studies provide a theoretical foundation for investigating the reflection and transmission characteristics of inhomogeneous waves in viscoelastic media. Since inhomogeneous waves account for attenuation angles, the reflection and transmission behavior of seismic waves at viscoelastic interfaces becomes considerably more complicated (Moradi and Innanen, 2016). Nevertheless, in practical seismic attenuation inversion, the influence of attenuation angles is often neglected, and only homogeneous waves are typically considered.

For attenuation inversion based on homogeneous waves, there are currently two main approaches for constructing linearized viscoelastic PP-wave reflection coefficients. The first approach incorporates phase velocity models containing $Q^{-1}$ into the frequency-extended Zoeppritz equation or their approximations to establish the relationship between frequency-dependent PP-wave reflection coefficients and seismic attenuation (Li and Liu, 2019; Chen, 2020a; Chen et al., 2022; Lan et al., 2024a, 2024b). This type of method is relatively simple in form and can be easily integrated into existing AVO or FDAVO inversion frameworks. The second approach is based on continuum mechanics and scattering theory, in which the viscoelastic stiffness matrix is expressed explicitly as a function of $Q$ and further decomposed into the sum of a homogeneous background term, an elastic perturbation term, and a viscoelastic perturbation term. By applying the Born integral and stationary phase approximation (Shaw and Sen, 2004), frequency-dependent linear PP-wave reflection coefficients are derived as functions of elastic parameters and seismic attenuation (Zong et al., 2015; Chen et al., 2018; Chen, 2020b). Compared with the first approach, these methods can more explicitly characterize the influence of seismic attenuation on reflection responses and provide a theoretical basis for viscoelastic seismic reflection analysis. Although existing seismic attenuation inversion methods have achieved important progress, most of them are mainly based on specific constant $Q$ models or frequency-dependent parameterizations and do not explicitly characterize the causal relationship between seismic dispersion and attenuation.

In linear viscoelastic media, seismic dispersion and attenuation satisfy the Kramers-Kronig relations (KKR, O'Donnell et al., 1981; Carcione, 2019), which reflects the causal constraint between the storage modulus and the dissipation modulus. Therefore, constructing frequency-dependent reflection coefficients that explicitly include seismic attenuation based on the KKR may further strengthen the physical relationship between seismic reflection dispersion and seismic attenuation and provide a new theoretical basis for seismic attenuation inversion. Motivated by this idea, this study starts from the KKR, expresses the frequency-dependent viscoelastic stiffness matrix as a function of seismic attenuation, and derives a frequency-dependent PP-wave reflection coefficient explicitly containing seismic attenuation based on scattering theory. On this basis, P- and S-wave attenuation are estimated using a time-frequency domain Bayesian inversion based on the derived reflection coefficient equation. To improve the resolution and stability of the inversion, we adopt a recursive frequency strategy with half-power overlapping Gaussian windows, in which the inversion result at the previous frequency serves as the prior model for the next frequency, and the prior models are built from well-log rock-physics modeling to provide low-frequency constraints. Synthetic and field seismic data are used to validate the effectiveness of the proposed method for estimating P- and S-wave attenuation and for gas reservoir prediction.

This paper is organized as follows. First, we present the derivation of the frequency-dependent PP-wave reflection coefficient equation based on the KKR and describe the time-frequency domain Bayesian inversion method for estimating P- and S-wave attenuation based on this equation. Next, synthetic and field seismic data are used to validate the effectiveness of the proposed method for predicting gas reservoirs. Finally, the applicability and limitations of the proposed method are discussed.

# 2 Theory and methodology

## 2.1 Derivation of the frequency-dependent PP-wave reflection coefficient equation using Kramers-Kronig relations

For linear viscoelastic media, the frequency-dependent complex modulus can be expressed as (Carcione, 2022)

$$\tilde{\chi}(\omega)=\chi_R(\omega)+i\chi_I(\omega), \tag{1}$$

where $\chi_R(\omega)$ and $\chi_I(\omega)$ denote the storage modulus and loss modulus, respectively, both are real valued and frequency dependent. $\omega=2\pi f$ is angular frequency, $f$ is frequency.

The quality factor $Q(\omega)$ is defined as the ratio between the storage modulus and loss modulus, and the inverse quality factor (or attenuation) $Q^{-1}(\omega)$ is defined as the reciprocal of the quality factor (Carcione, 2022)

$$Q(\omega)=\frac{\mathrm{Re}\left[\tilde{\chi}(\omega)\right]}{\mathrm{Im}\left[\tilde{\chi}(\omega)\right]}=\frac{\chi_R(\omega)}{\chi_I(\omega)},\ Q^{-1}(\omega)=\frac{1}{Q(\omega)}. \tag{2}$$

For linear viscoelastic media, physical causality (the stress-strain response of the system must occur after the externally applied stress) requires that the storage modulus $\chi_R(\omega)$ and loss modulus $\chi_I(\omega)$ satisfy the Kramers-Kronig relations (KKRs) (Carcione et al, 2019)

$$\begin{aligned}\chi_R(\omega)-\chi_R(\infty)&=\left(\frac{1}{\pi\omega}\right)*\chi_I(\omega)=\frac{1}{\pi}\mathcal{P}\int_{-\infty}^{\infty}\frac{\chi_I(\omega')}{\omega-\omega'}d\omega'\\ \chi_I(\omega)&=-\left(\frac{1}{\pi\omega}\right)*\left[\chi_R(\omega)-\chi_R(\infty)\right]=-\frac{1}{\pi}\mathcal{P}\int_{-\infty}^{\infty}\frac{\chi_R(\omega')}{\omega-\omega'}d\omega'\end{aligned}, \tag{3}$$

where $\mathcal{P}$ denotes the Cauchy principal value, $\omega$ and $\omega'$ are angular frequencies.

Once one modulus is known over the entire frequency range, the other can be obtained using this exact KKRs. However, in practice, only the real part of the modulus is usually available over a limited frequency range. When modulus dispersion varies slowly with frequency, the KKR can be approximated locally as (O'Donnell et al., 1981; Christensen, 1982; Mikhaltsevitch et al., 2016)

$$\chi_R(\omega)\approx\chi_R(\omega_0)\left[1+\frac{2}{\pi}\int_{\ln\omega_0}^{\ln\omega}Q^{-1}(\omega')d(\ln\omega')\right], \tag{4}$$

where $\omega_0=2\pi f_0$ is reference angular frequency, $f_0$ is reference frequency.

Eq. 4 are applicable to arbitrary elastic moduli, such as the bulk, shear, and Young's moduli (Gurevich and Carcione, 2022). In broadband rock physics measurements, the local KKR in Eq. 4 is used to demonstrate the causal relationship between modulus dispersion and modulus attenuation (Mikhaltsevitch et al., 2014, 2016; Chapman et al., 2016; Yin et al., 2017; Zhao et al., 2021). The measured modulus dispersion agrees well with that predicted from the local KKR, providing evidence for the validity of the local approximation KKR.

Eq. 4 establishes an analytical relationship between the storage modulus and attenuation, which can be used to construct a frequency-dependent stiffness matrix incorporating attenuation. Furthermore, scattering theory is employed to derive a linearized reflection coefficient equation for viscoelastic media. Replacing $\tilde{\chi}(\omega)$ with the P-wave modulus $\tilde{M}(\omega)=M_R(\omega)+iM_I(\omega)$ and shear modulus $\tilde{\mu}(\omega)=\mu_R(\omega)+i\mu_I(\omega)$, respectively, the local approximate KKR for the P-wave and shear moduli are

$$\begin{aligned}\tilde{M}(\omega) &\approx M_R(\omega_0)\left[1+\frac{2}{\pi}\int_{\ln\omega_0}^{\ln\omega} Q_P^{-1}(\omega')d(\ln\omega')\right] \\ &\quad +iM_R(\omega_0)\left[1+\frac{2}{\pi}\int_{\ln\omega_0}^{\ln\omega} Q_P^{-1}(\omega')d(\ln\omega')\right]Q_P^{-1}(\omega) \\ \tilde{\mu}(\omega) &\approx \mu_R(\omega_0)\left[1+\frac{2}{\pi}\int_{\ln\omega_0}^{\ln\omega} Q_S^{-1}(\omega')d(\ln\omega')\right] \\ &\quad +i\mu_R(\omega_0)\left[1+\frac{2}{\pi}\int_{\ln\omega_0}^{\ln\omega} Q_S^{-1}(\omega')d(\ln\omega')\right]Q_S^{-1}(\omega)\end{aligned}, \quad (5)$$

where $Q_P^{-1}(\omega)$ and $Q_S^{-1}(\omega)$ denote P-wave and S-wave attenuation, respectively.

For isotropic viscoelastic media, the perturbation stiffness matrix $\Delta\mathbf{C}(\omega)$ is expressed as (Chen et al, 2018)

$$\Delta\tilde{\mathbf{C}}(\omega)=\begin{bmatrix}\Delta\tilde{C}_{33}(\omega) & \Delta\tilde{C}_{12}(\omega) & \Delta\tilde{C}_{12}(\omega) & 0 & 0 & 0\\ \Delta\tilde{C}_{12}(\omega) & \Delta\tilde{C}_{33}(\omega) & \Delta\tilde{C}_{12}(\omega) & 0 & 0 & 0\\ \Delta\tilde{C}_{12}(\omega) & \Delta\tilde{C}_{12}(\omega) & \Delta\tilde{C}_{33}(\omega) & 0 & 0 & 0\\ 0 & 0 & 0 & \Delta\tilde{C}_{55}(\omega) & 0 & 0\\ 0 & 0 & 0 & 0 & \Delta\tilde{C}_{55}(\omega) & 0\\ 0 & 0 & 0 & 0 & 0 & \Delta\tilde{C}_{55}(\omega)\end{bmatrix}, \quad (6)$$

where $\Delta\tilde{C}_{12}(\omega)=\Delta\tilde{C}_{33}(\omega)-2\Delta\tilde{C}_{55}(\omega)$, $\Delta\tilde{C}_{33}(\omega)=\Delta\tilde{M}(\omega)$, $\Delta\tilde{C}_{55}(\omega)=\Delta\tilde{\mu}_R(\omega)$.

According to the Eq. 5, under the assumptions of weak contrasts in stiffness coefficients across the interface and weak attenuation, the perturbations of $\tilde{C}_{33}(\omega)$ and $\tilde{C}_{55}(\omega)$ can be expressed as follows (the detailed derivations is provided in Appendix A)

$$\Delta\tilde{C}_{33}(\omega) \approx \Delta M_R(\omega_0) + M_R(\omega_0)\frac{2}{\pi}\int_{\ln\omega_0}^{\ln\omega} \Delta Q_P^{-1}(\omega')d(\ln\omega')$$
$$+i\left\{M_R(\omega_0)\Delta Q_P^{-1}(\omega) + \frac{2}{\pi}M_R(\omega_0)\left[\Delta Q_P^{-1}(\omega)\int_{\ln\omega_0}^{\ln\omega} Q_P^{-1}(\omega')d(\ln\omega') + Q_P^{-1}(\omega)\int_{\ln\omega_0}^{\ln\omega} \Delta Q_P^{-1}(\omega')d(\ln\omega')\right]\right\}$$
$$\Delta\tilde{C}_{55}(\omega) \approx \Delta\mu_R(\omega_0) + \mu(\omega_0)\frac{2}{\pi}\int_{\ln\omega_0}^{\ln\omega} \Delta Q_S^{-1}(\omega')d(\ln\omega')$$
$$+i\left\{\mu_R(\omega_0)\Delta Q_S^{-1}(\omega) + \frac{2}{\pi}\mu_R(\omega_0)\left[\Delta Q_S^{-1}(\omega)\int_{\ln\omega_0}^{\ln\omega} Q_S^{-1}(\omega')d(\ln\omega') + Q_S^{-1}(\omega)\int_{\ln\omega_0}^{\ln\omega} \Delta Q_S^{-1}(\omega')d(\ln\omega')\right]\right\}$$
.(7)

Using the Born integral and the stationary phase approximation, Shaw and Sen (2004) established a theoretical framework for calculating linearized PP-wave reflection coefficients in anisotropic media. The linearized PP-wave reflection coefficient is related to the scattering function $\Re$ by

$$R_{PP} = \frac{1}{4\rho\cos^2\theta}\Re. \quad (8)$$

For isotropic viscoelastic media, the scattering function $\Re$ can be write as (Zong et al., 2015; Chen et al., 2018)

$$\Re = \Delta\rho\cos 2\theta + \frac{\rho}{M}\left(\Delta\tilde{C}_{33}(\omega) - 2\Delta\tilde{C}_{55}(\omega)\sin^2 2\theta\right), \quad (9)$$

where $M = M_R(\omega_0)$ and $\rho$ denote the P-wave modulus and density of the background medium, respectively, and $\theta$ is the incident angle.

Using Eqs. 7-9 the following frequency-dependent complex reflection coefficient can be obtained

$$R_{PP}(\theta,\omega)=\frac{1}{4\cos^2\theta}\frac{\Delta M_R(\omega_0)}{M_R(\omega_0)}-2\sin^2\theta\frac{\mu_R(\omega_0)}{M_R(\omega_0)}\frac{\Delta\mu_R(\omega_0)}{\mu_R(\omega_0)}+\frac{\cos 2\theta}{4\cos^2\theta}\frac{\Delta\rho}{\rho}$$

$$+\frac{1}{2\pi\cos^2\theta}\int_{\ln\omega_0}^{\ln\omega}\Delta Q_P^{-1}(\omega')d(\ln\omega')-\frac{4\sin^2\theta}{\pi}\frac{\mu_R(\omega_0)}{M_R(\omega_0)}\int_{\ln\omega_0}^{\ln\omega}\Delta Q_S^{-1}(\omega')d(\ln\omega')$$

$$+i\left\{\begin{array}{l}\frac{1}{4\cos^2\theta}\Delta Q_P^{-1}(\omega)-2\sin^2\theta\frac{\mu_R(\omega_0)}{M_R(\omega_0)}\Delta Q_S^{-1}(\omega)\\+\frac{1}{2\pi\cos^2\theta}\left[\Delta Q_P^{-1}(\omega)\int_{\ln\omega_0}^{\ln\omega}Q_P^{-1}(\omega')d(\ln\omega')+Q_P^{-1}(\omega)\int_{\ln\omega_0}^{\ln\omega}\Delta Q_P^{-1}(\omega')d(\ln\omega')\right]\\-\frac{4\sin^2\theta}{\pi}\frac{\mu_R(\omega_0)}{M_R(\omega_0)}\left[\Delta Q_S^{-1}(\omega)\int_{\ln\omega_0}^{\ln\omega}Q_S^{-1}(\omega')d(\ln\omega')+Q_S^{-1}(\omega)\int_{\ln\omega_0}^{\ln\omega}\Delta Q_S^{-1}(\omega')d(\ln\omega')\right]\end{array}\right\}.$$

(10)

Eq. 10 indicates that, in the linear viscoelastic medium, the reflection coefficient at a given frequency is determined by seismic attenuation across the range from the reference to the target frequency. However, the real and imaginary parts of the reflection coefficient involve integrals of the unknown P- and S-wave attenuation at different frequencies, which limits its application in seismic inversion. According to the numerical analysis in Appendix B, the imaginary part of the Eq. 10 can be neglected during inversion under weak seismic attenuation, and Eq. 10 is simplified to

$$R_{PP}(\theta,\omega)=\frac{1}{4\cos^2\theta}\frac{\Delta M_R(\omega_0)}{M_R(\omega_0)}-2\sin^2\theta\frac{\mu_R(\omega_0)}{M_R(\omega_0)}\frac{\Delta\mu_R(\omega_0)}{\mu_R(\omega_0)}+\frac{\cos 2\theta}{4\cos^2\theta}\frac{\Delta\rho}{\rho}$$

$$+\frac{1}{2\pi\cos^2\theta}\int_{\ln\omega_0}^{\ln\omega}\Delta Q_P^{-1}(\omega')d(\ln\omega')-\frac{4\sin^2\theta}{\pi}\frac{\mu_R(\omega_0)}{M_R(\omega_0)}\int_{\ln\omega_0}^{\ln\omega}\Delta Q_S^{-1}(\omega')d(\ln\omega')$$

.(11)

Eq. 11 accounts for the effects of frequency-dependent P-wave and S-wave attenuation on the reflection coefficient. There commonly exist multiscale and multitype heterogeneities in subsurface rocks, including variations in lithology, permeability, and fluid properties (Mukerji, 1995; Bailly et al., 2019; Zhao et al., 2023).

Different heterogeneities possess different attenuation characteristic frequencies, and their coupled dispersion-attenuation effects jointly control the overall seismic attenuation behavior (Zhao et al., 2021; Guo and Gurevich, 2020; Wang et al., 2026), often causing attenuation to behave approximately as a constant within the seismic frequency band (Moradi et al., 2014; Mavko et al., 2020). Based on this consideration, we assume that P-wave and S-wave attenuation remain approximately constant near the reference angular frequency $\omega_0$. Therefore, Eq. 11 further becomes

$$\begin{aligned} R_{PP}(\theta,\omega) &= \frac{1}{4\cos^2\theta}\frac{\Delta M_R(\omega_0)}{M_R(\omega_0)} - 2\sin^2\theta\frac{\mu_R(\omega_0)}{M_R(\omega_0)}\frac{\Delta\mu_R(\omega_0)}{\mu_R(\omega_0)} + \frac{\cos 2\theta}{4\cos^2\theta}\frac{\Delta\rho}{\rho} \\ &+ \frac{1}{2\pi\cos^2\theta}\ln\frac{\omega}{\omega_0}\Delta Q_P^{-1}(\omega_0) - \frac{4\sin^2\theta}{\pi}\frac{\mu_R(\omega_0)}{M_R(\omega_0)}\ln\frac{\omega}{\omega_0}\Delta Q_S^{-1}(\omega_0) \end{aligned}. \quad (12)$$

When $\omega = \omega_0$, Eq. 12 reduces to the reflection coefficient at the reference angular frequency

$$R_{PP}(\theta) = \frac{1}{4\cos^2\theta}\frac{\Delta M_R(\omega_0)}{M_R(\omega_0)} - 2\sin^2\theta\frac{\mu_R(\omega_0)}{M_R(\omega_0)}\frac{\Delta\mu_R(\omega_0)}{\mu_R(\omega_0)} + \frac{\cos 2\theta}{4\cos^2\theta}\frac{\Delta\rho}{\rho}. \quad (13)$$

Eq. 13 is equivalent to the elastic P-wave reflection approximation proposed by Fatti et al. (1994), which is widely used in seismic inversion for P- and S-wave impedance. Therefore, the elastic reflection term in Eq. 12 is replaced by Fatti's reflection coefficient, and Eq. 12 is rewritten as:

$$\begin{aligned} R_{PP}(\theta,\omega) &= \frac{1}{2}\left(1+\tan^2\theta\right)\frac{\Delta Z_p(\omega_0)}{Z_p(\omega_0)} - 4\sin^2\theta\frac{Z_s^2(\omega_0)}{Z_p^2(\omega_0)}\frac{\Delta Z_s(\omega_0)}{Z_s(\omega_0)} - \left[\frac{1}{2}\tan^2\theta - 2\sin^2\theta\frac{Z_s^2(\omega_0)}{Z_p^2(\omega_0)}\right]\frac{\Delta\rho}{\rho} \\ &+ \frac{1}{2\pi\cos^2\theta}\ln\frac{\omega}{\omega_0}\Delta Q_P^{-1}(\omega_0) - \frac{4\sin^2\theta}{\pi}\frac{Z_s^2(\omega_0)}{Z_p^2(\omega_0)}\ln\frac{\omega}{\omega_0}\Delta Q_S^{-1}(\omega_0) \end{aligned}, \quad (14)$$

where $Z_p(\omega_0)$ and $Z_s(\omega_0)$ are the P- and S-wave impedances at the reference angular frequency $\omega_0$.

Eq. 14 serves as the key equation for the subsequent seismic attenuation inversion. Due to seismic wave energy is mainly concentrated around the dominant frequency, the phase velocity at the dominant frequency plays a dominant role in seismic propagation for weakly attenuating viscoelastic media (Li and Liu, 2019). Based on this consideration, the seismic dominant frequency is selected as the reference frequency $f_0$ in this study.

**2.2 Estimation of P- and S-wave attenuations from well logs**

Because seismic data lack low-frequency components, initial model constraints are necessary to compensate for the low-frequency components of the inversion parameters and to reduce the non-uniqueness of the inversion. Typically, the initial model is obtained through low-frequency filtering of well logs and spatial interpolation. Acoustic logging provides the P- and S-wave velocities and density of the formation, but it cannot directly measure P- and S-wave attenuation (Mavko et al., 2005). However, P- and S-wave attenuation can be estimated from conventional well-log data through rock-physics modeling (Mavko et al., 2005; Dvorkin and Mavko, 2006; Teng et al., 2012). In this study, we use the Dvorkin-Mavko attenuation model (Mavko et al., 2020) to estimate P- and S-wave attenuation from well logs and thereby establish the initial model of P- and S-wave attenuation for subsequent seismic inversion. The specific formulas are given in Appendix C.

Estimation of P- and S-wave attenuation involves several unknown parameters associated with the solid and fluid phases. Among these, the dry rock moduli are relatively difficult to obtain and are currently usually derived through rock physics modeling. The critical porosity model proposed by Nur (1998) is a practical model that has been widely used in rock physics modeling (Markov et al., 2013; Zhang et al., 2021; Ba et al., 2023; Luo et al., 2025). This model relates porosity, matrix moduli, and dry rock moduli through the critical porosity:

$$K_{dry} = K_m \left(1 - \frac{\phi}{\phi_c}\right), \tag{15a}$$

$$G_{dry} = G_m \left(1 - \frac{\phi}{\phi_c}\right), \tag{15b}$$

where $\phi$ is porosity, $\phi_c$ is the critical porosity, and $K_{dry}$ and $G_{dry}$ are the bulk and shear moduli of the dry rock. $K_m$ and $G_m$ are the bulk and shear moduli of the matrix, obtained from the Voigt-Reuss-ill (V-R-H) average (Hill, 1952):

$$K_m = \frac{1}{2}\left[\sum_{i=1}^{M} K_i \varphi_i + \left(\sum_{i=1}^{M} K_i / \varphi_i\right)^{-1}\right], \tag{16a}$$

$$G_m = \frac{1}{2}\left[\sum_{i=1}^{M} G_i \varphi_i + \left(\sum_{i=1}^{M} G_i / \varphi_i\right)^{-1}\right], \tag{16b}$$

where $\varphi_i$, $K_i$ and $G_i$ represent the fraction, bulk modulus, and shear modulus of the *i*-th mineral component, respectively; and M is the total number of mineral components.

For subsurface media, rocks at different depths may have different critical porosities owing to gradual changes in lithology and lithofacies. Therefore, a variable critical porosity model can more reasonably describe the pore structures of complex reservoirs (Zhang et al., 2021; Ba et al., 2023). To this end, we treat the critical porosity as an unknown parameter and estimate it through an optimization method using the P- and S-wave velocities of well as observed data. Rock physics parameter inversion usually uses the P-wave velocity predicted by the Gassmann-Wood equation (Eq. C2) in the low-frequency limit to calibrate the log P-wave velocity. However, the logging frequency is on the order of kHz, which is far above the validity range of the Gassmann-Wood equation. We therefore use the P-wave velocity predicted by the Gassmann-Hill equation (Eq. C3) in the high-frequency limit to calibrate the log P-wave velocity and construct the following objective function for critical porosity inversion:

$$J(\phi_c) = \min \frac{1}{2} \left\| V_{p_pre}(\varphi, \phi, \phi_c) - V_{p_obs} \right\|_2^2 + \frac{1}{2} \left\| V_{s_pre}(\varphi, \phi, \phi_c) - V_{s_obs} \right\|_2^2, \tag{17}$$

where $V_{p_obs}$ and $V_{s_obs}$ are the observed P- and S-wave velocities from well logs, and $V_{p_pre}$ and $V_{s_pre}$ are the P- and S-wave velocities predicted by rock physics modeling. Because of the nonlinear relationship between $\phi_c$ and $V_{p_pre}$, we use the particle swarm optimization algorithm to solve Eq. 17. After obtaining $\phi_c$, the log derived P- and S-wave attenuation can be estimated by combining Eq. 15 with the equations in Appendix C. The workflow for the above estimation of seismic attenuation from well logs is shown in Fig. 1.

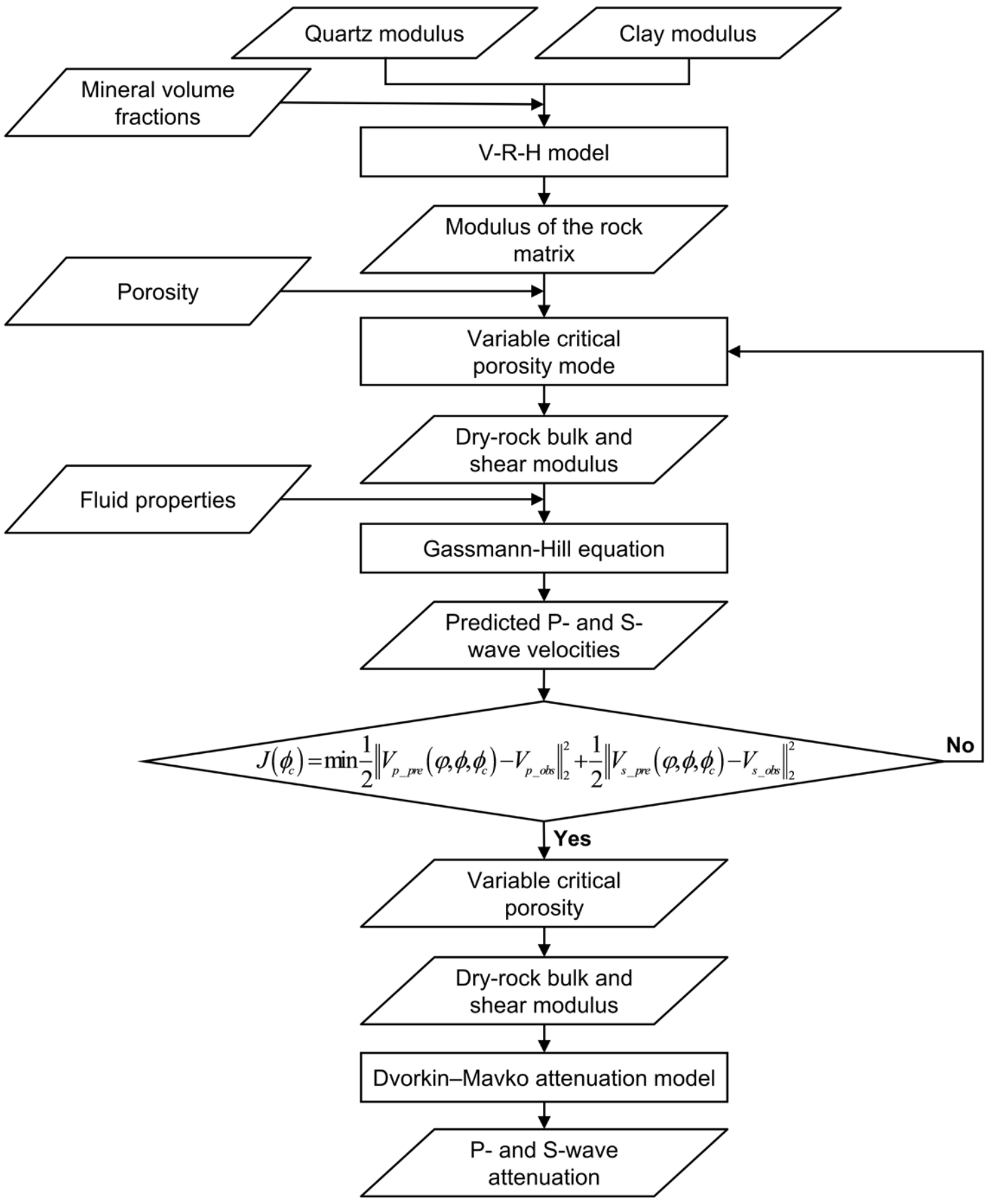

Fig.1. Workflow for calculating P- and S-wave attenuation from well logs.

## 2.3 Seismic P- and S-wave attenuation inversion in the time-frequency domain

According to the exploding reflector principle, the seismic wavefield is equal to the superposition of the frequency-dependent reflected wavefields from each interface. The frequency-domain forward modeling of the angle-dependent seismic wavefield is as follows

$$s(\theta,t)=\mathcal{F}^{-1}_{\omega\to t}\left[\sum_{i=1}^{N}W(\theta,\omega)R_{PP}(\theta,\omega,\tau_i)e^{-i\omega\tau_i}\right], \tag{18}$$

where $W(\theta,\omega)$ is the frequency spectrum of the seismic wavelets at different angles, $\mathcal{F}^{-1}$ denotes the inverse Fourier transform, $t$ is time, and $N$ is the number of time samples. $R_{PP}(\theta,\omega,\tau_i)$ is the frequency-dependent reflection coefficient at time $\tau_i$.

According to Buland and More (2003), $R_{PP}(\theta,\omega,t)$ can be written as

$$
\begin{aligned}
R_{PP}(\theta,\omega,t) &= \frac{1}{2}\left(1+\tan^2\theta\right)\frac{\partial \ln Z_p(\omega_0,t)}{\partial t} \\
&-4\sin^2\theta\frac{Z_s^2(\omega_0)}{Z_p^2(\omega_0)}\frac{\partial \ln Z_s(\omega_0,t)}{\partial t}-\left[\frac{1}{2}\tan^2\theta-2\sin^2\theta\frac{Z_s^2(\omega_0)}{Z_p^2(\omega_0)}\right]\frac{\partial \ln\rho(t)}{\partial t}. \quad (19) \\
&+\frac{1}{2\pi\cos^2\theta}\ln\frac{\omega}{\omega_0}\frac{\partial Q_P^{-1}(\omega_0,t)}{\partial t}-\frac{4\sin^2\theta}{\pi}\frac{Z_s^2(\omega_0)}{Z_p^2(\omega_0)}\ln\frac{\omega}{\omega_0}\frac{\partial Q_S^{-1}(\omega_0,t)}{\partial t}
\end{aligned}
$$

Time-frequency analysis can extract multi-scale information from seismic data and characterize the time-varying amplitude and phase of different frequency components of seismic data, making it a key method for extracting seismic wave dispersion and attenuation information (Castagna et al., 2003; Goloshubin et al., 2006; Wilson et al., 2009; Chen et al., 2012). Compared with time-domain inversion, inversion in the time-frequency domain can exploit the multi-scale characteristics of the seismic time-frequency spectrum and fully utilize the roles of different frequency components of seismic signals in inversion, thereby obtaining higher-precision inversion results (Wang et al., 2026). Therefore, the inversion is performed in the time-frequency domain in this paper.

Among numerous time-frequency analysis methods, windowed time-frequency transforms represented by the short-time Fourier transform (Gabor, 1946) are widely used because of their linearity and high computational efficiency (Stockwell, 1999; Pinnegar and Mansinha, 2003; Chen et al., 2009; Gholami, 2013; Li et al., 2016; Wang, 2021; Pegah et al., 2025; Wang et al., 2025). The windowed time-frequency transform of the $s(\theta,t)$ is as follows

$$S(\theta,f,t)=\int_{-\infty}^{+\infty} s(\theta,t)\,g(t-\varsigma,f)\,e^{-i2\pi f\varsigma}d\tau, \quad (20)$$

where $S(\theta, f, t)$ is the time-frequency spectrum of $s(\theta, t)$, $\varsigma$ is the integration time, $g(t, f)$ is the Gaussian window function, different window functions correspond to different time-frequency transforms. In this study, the seismic time-frequency spectrum of seismic is computed by the adaptive generalized S-transform (Wang et al., 2025).

Eq. 20 can be expressed in the following convolution form

$$S(\theta, f, t) = g(t, f) * \left[ s(\theta, t) e^{-i2\pi f \varsigma} \right]. \tag{21}$$

Substituting Eq. 18 into Eq. 21 yields the time-frequency representation of the frequency-dependent pre-stack seismic

$$S(\theta, f, t) = g(t, f) * \left\{ \mathcal{F}_{f \to t}^{-1} \left[ \sum_{i=1}^{N} W(\theta, f) R_{PP}(\theta, \omega, \tau_i) e^{-i2\pi f \tau_i} \right] e^{-i2\pi f \varsigma} \right\}. \tag{22}$$

At the *j*-th angle $\theta_j$ and the *k*-th frequency $f_k$, Eq. 22 can be expressed in the following matrix form

$$\mathbf{S}_{k,j} = \mathbf{G}_k \mathbf{M}_k \mathbf{F}^{-1} \mathbf{W}_j \mathbf{\Lambda}_j \left( \mathbf{I}_5 \otimes \mathbf{P} \right) \left( \mathbf{I}_5 \otimes \mathbf{D}_0 \right) \mathbf{m}, \tag{23}$$

where $\mathbf{S}_{k,j} \in \mathbb{C}^{N_t \times 1}$ is the seismic wavefield at the *j*-th angle and *k*-th frequency, $\mathbf{F}^{-1} \in \mathbb{C}^{N_t \times N_f}$ is the inverse Fourier transform matrix, $\mathbf{G}_k \in \mathbb{R}^{N_t \times N_t}$ and $\mathbf{M}_k \in \mathbb{C}^{N_t \times N_t}$ are the window matrix and modulation matrix at the *k*-th frequency, $\mathbf{W}_j \in \mathbb{C}^{N_f \times N_f}$ and $\mathbf{\Lambda}_j \in \mathbb{R}^{N_f \times 5N_f}$ are the wavelet spectrum matrix and coefficient matrix at the *j*-th angle, $\mathbf{P} \in \mathbb{C}^{N_f \times N_t}$ is the phase shift matrix, $\mathbf{D}_0 \in \mathbb{R}^{N_t \times (N_t+1)}$ is the difference matrix, $\mathbf{I}_5 \in \mathbb{R}^{5 \times 5}$ is the identity matrix, $\mathbf{m} \in \mathbb{R}^{5(N_t+1) \times 1}$ is the parameter vector to be inverted, and $\otimes$ denotes the Hadamard product. The specific forms of the above matrices are presented in Appendix D.

Extending Eq.23 to $M$ angles yields

$$\underbrace{\begin{bmatrix} \mathbf{S}_{k,1} \\ \mathbf{S}_{k,2} \\ \vdots \\ \mathbf{S}_{k,M} \end{bmatrix}}_{\hat{\mathbf{d}}_k \in \mathbb{C}^{MN_t \times 1}} = \underbrace{\begin{bmatrix} \mathbf{G}_k \mathbf{M}_k \mathbf{F}^{-1} \mathbf{W}_1 \mathbf{\Lambda}_1 (\mathbf{I}_5 \otimes \mathbf{P})(\mathbf{I}_5 \otimes \mathbf{D}_0) \\ \mathbf{G}_k \mathbf{M}_k \mathbf{F}^{-1} \mathbf{W}_2 \mathbf{\Lambda}_2 (\mathbf{I}_5 \otimes \mathbf{P})(\mathbf{I}_5 \otimes \mathbf{D}_0) \\ \vdots \\ \mathbf{G}_k \mathbf{M}_k \mathbf{F}^{-1} \mathbf{W}_M \mathbf{\Lambda}_M (\mathbf{I}_5 \otimes \mathbf{P})(\mathbf{I}_5 \otimes \mathbf{D}_0) \end{bmatrix}}_{\hat{\mathbf{\Psi}}_k \in \mathbb{C}^{MN_t \times 5(N_t+1)}} \mathbf{m}. \tag{24}$$

Separating the real and imaginary parts of Eq. 24 yields

$$\underbrace{\begin{bmatrix} \mathrm{Re}(\hat{\mathbf{d}}_k) \\ \mathrm{Im}(\hat{\mathbf{d}}_k) \end{bmatrix}}_{\mathbf{d}_k} = \underbrace{\begin{bmatrix} \mathrm{Re}(\hat{\mathbf{\Psi}}_k) \\ \mathrm{Im}(\hat{\mathbf{\Psi}}_k) \end{bmatrix}}_{\mathbf{\Psi}_k} \mathbf{m}, \tag{25}$$

where $\mathrm{Re}(\bullet)$ and $\mathrm{Im}(\bullet)$ denote the real and imaginary part operators, respectively.

Adding noise $\mathbf{e}$ to Eq. 25 yields the following time-frequency domain frequency-dependent seismic forward modeling equation at the $k$-th frequency $f_k$

$$\mathbf{d}_k = \mathbf{\Psi}_k \mathbf{m} + \mathbf{e} \tag{26}$$

Under the Bayesian inversion framework, Eq. 25 is solved. Assuming that the noise follows a Gaussian distribution with mean 0 and covariance matrix $\mathbf{C}_\mathbf{d}$, and that the prior distribution of model $\mathbf{m}$ follows a Gaussian distribution with mean $\mathbf{m}_0$ and covariance matrix $\mathbf{C}_\mathbf{m}$, the posterior distribution of model $\mathbf{m}$ for solving Eq. 26 is (Buland and More, 2003; Grana et al., 2024)

$$P(\mathbf{m} \mid \mathbf{d}_k) = \left\{ -\frac{1}{2} (\mathbf{d}_k - \mathbf{\Psi}_k \mathbf{m})^T \mathbf{C}_\mathbf{d} (\mathbf{d}_k - \mathbf{\Psi}_k \mathbf{m}) - \frac{1}{2} (\mathbf{m} - \mathbf{m}_0)^T \mathbf{C}_\mathbf{m} (\mathbf{m} - \mathbf{m}_0) \right\}. \tag{27}$$

The posterior mean of model $\mathbf{m}$ is

$$\boldsymbol{\mu}_{\mathbf{m}|\mathbf{d}_k} = \mathbf{m}_0 + \mathbf{C}_\mathbf{m} \mathbf{\Psi}_k^T \left( \mathbf{\Psi}_k \mathbf{C}_\mathbf{m} \mathbf{\Psi}_k^T + \mathbf{C}_\mathbf{d} \right)^{-1} (\mathbf{d}_k - \mathbf{\Psi}_k \mathbf{m}_0). \tag{28}$$

Since the time-frequency window function is local, the $\mathbf{d}_k$ is a composite response of the local seismic spectrum centered at $f_k$. Therefore, the model obtained by inverting Eq. 28 can only recover information in the corresponding frequency band, and Eq. 28 should be rewritten as

$$\boldsymbol{\mu}_{\mathbf{m}_k|\mathbf{d}_k} = \mathbf{m}_0 + \mathbf{C}_\mathbf{m}\boldsymbol{\Psi}_k^T\left(\boldsymbol{\Psi}_k\mathbf{C}_\mathbf{m}\boldsymbol{\Psi}_k^T + \mathbf{C}_\mathbf{d}\right)^{-1}\left(\mathbf{d}_k - \boldsymbol{\Psi}_k\mathbf{m}_0\right). \tag{29}$$

The low-frequency time-frequency spectrum recovers the low-frequency information of the model, which characterizes the large-scale features of the formation. The high-frequency time-frequency spectrum recovers the high-frequency information of the model, which characterizes the small-scale details of the formation. On this basis, we use the inversion result at the previous frequency as the prior model for the next frequency. We traverse the entire seismic frequency band iteratively to obtain the final inversion model.

To ensure that structural information recovered at one scale is smoothly propagated to the next without introducing spectral gaps or excessive redundancy, the neighboring Gaussian windows should be selected to provide sufficient spectral overlap. Consequently, we employ a half-power overlap criterion, which requires the frequency-domain effective supports of adjacent windows to intersect at their half-power points, namely, $\left|G(f_i)\right|^2 = \left|G(f_{i+1})\right|^2 = \frac{1}{2}$. Here, $G(f_i)$ and $G(f_{i+1})$ represent the Gaussian windows corresponding to the *i*-th and (*i*+1)-th frequencies, respectively. Based on this condition, the implicit equation $f_{i+1} = f_i + \sqrt{\ln 2}\left(\delta_{f_i} + \delta_{f_{i+1}}\right)$ can be further derived, in which $\delta_{f_i}$ and $\delta_{f_{i+1}}$ are the standard deviation of $G(f_i)$ and $G(f_{i+1})$, respectively. Given that adjacent inversion frequencies are densely distributed, we apply the approximation $\delta_{f_i} \approx \delta_{f_{i+1}}$, which yields the recursive expression $f_{i+1} = f_i + 2\sqrt{\ln 2}\delta_{f_i}$. Beginning with the lowest inversion frequency, the inversion sequence is therefore constructed progressively until the maximum effective seismic frequency is reached. For a clearer illustration of the aforementioned inversion process, the corresponding workflow is provided in Fig. 2.

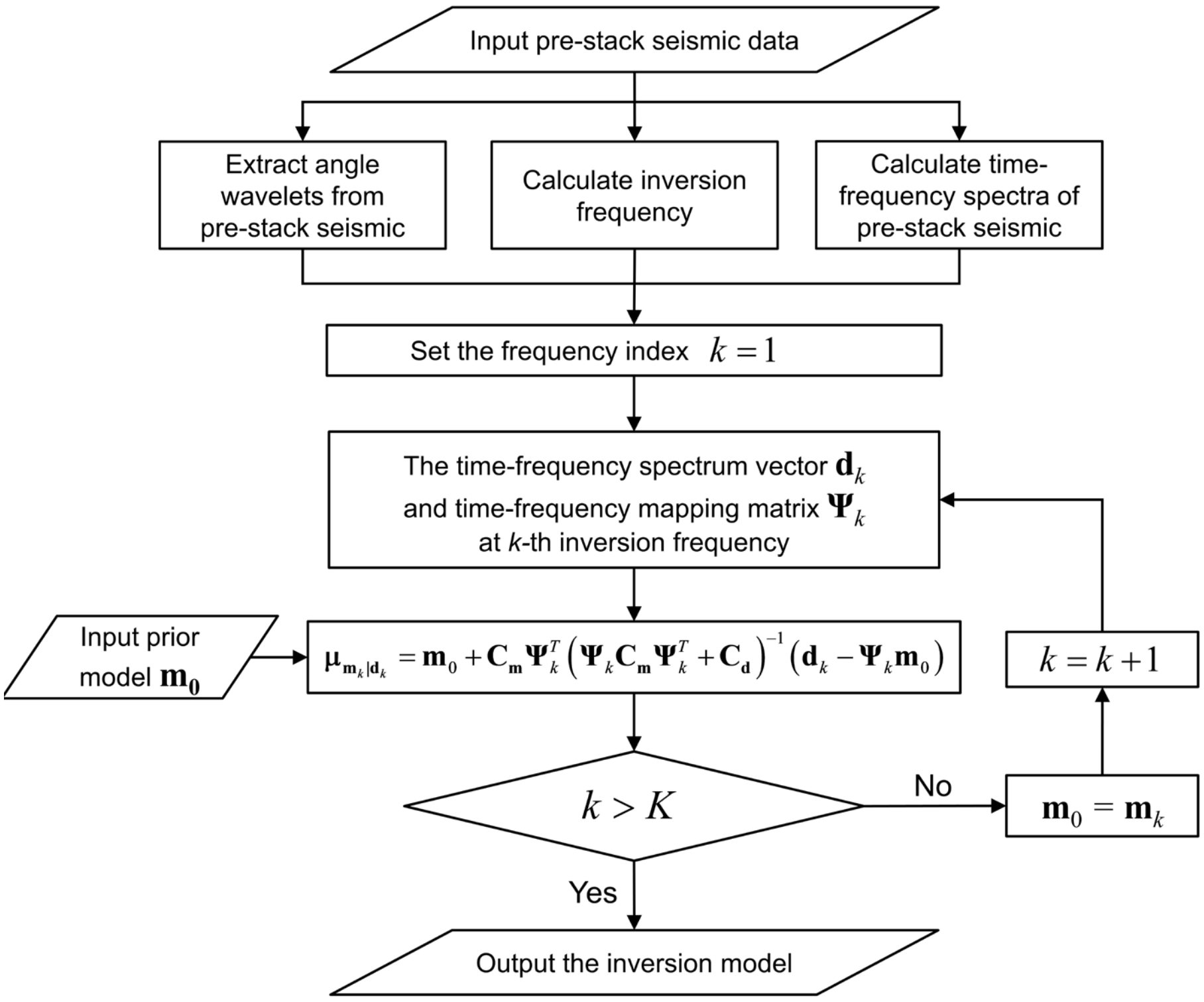

Fig.2. Workflow for seismic P- and S-wave attenuation inversion in the time-frequency domain.

# 3 Synthetic Data Test

In this section, synthetic seismic data are used to verify the validity of the proposed inversion method. Figure 3 shows the theoretical model. A Ricker wavelet with a dominant frequency of 30 Hz is adopted, and frequency-dependent seismic gathers at different angles, as shown in Figure 4, are obtained via frequency-domain forward modeling.

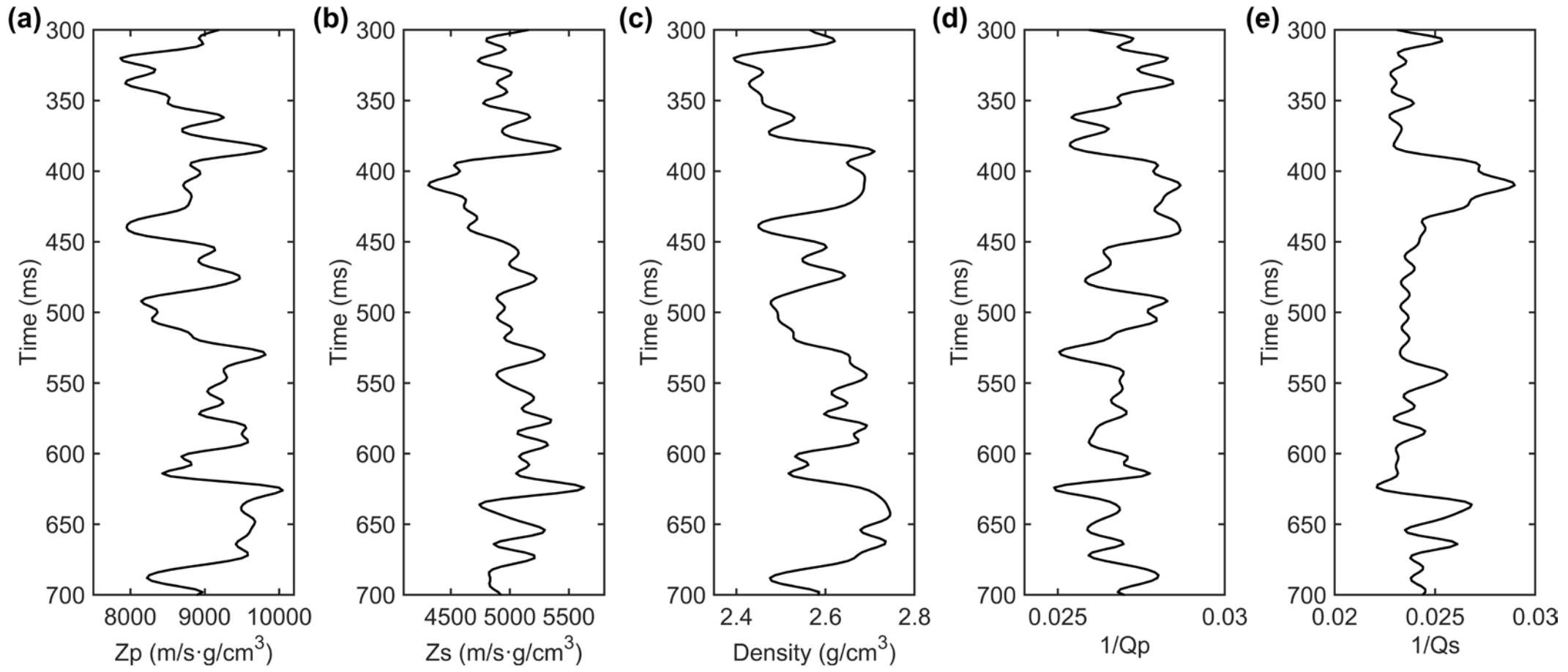

Fig. 3. Theoretical model. (a) P-wave impedance, (b) S-wave impedance, (c) density, (d) P-wave attenuation, and (e) S-wave attenuation.

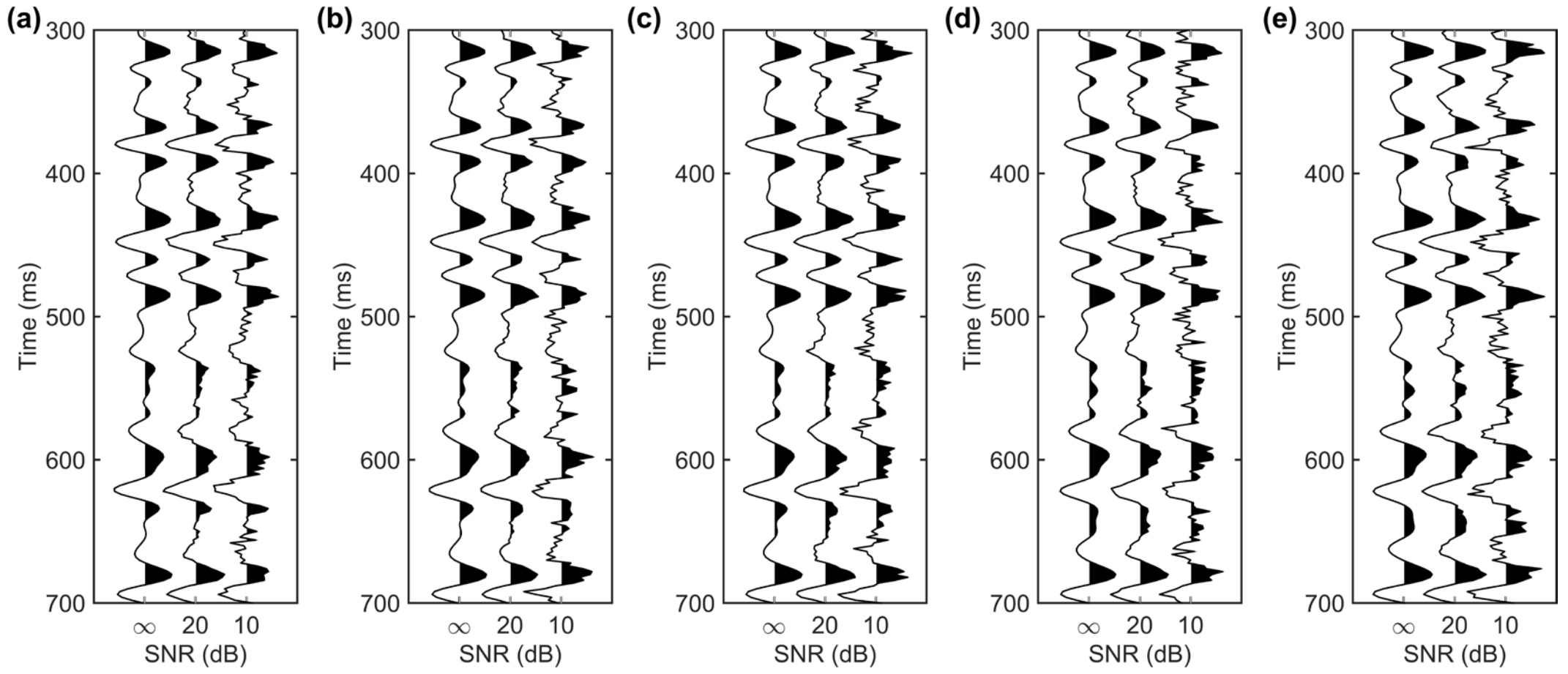

Fig. 4. Pre-stack synthetic seismic data. (a) $\theta = 5^{\circ}$, (b) $\theta = 10^{\circ}$, (c) $\theta = 15^{\circ}$, (d) $\theta = 25^{\circ}$, and (e) $\theta = 35^{\circ}$.

We first use the synthetic seismic signals with an signal-to-noise ratio (SNR) of 20 dB as an example to elaborate the implementation workflow of the inversion method. Fig. 5 shows the time-frequency amplitude spectra (magnitudes of the time-frequency spectra) at different angles for an SNR of 20 dB. In seismic inversion, the prior model $\mathbf{m}_0$ is generally obtained by low-pass filtering the true model. In this work, the cutoff frequency of the low-pass filter is set to 10 Hz, accordingly, the minimum inversion frequency is set to 15 Hz. The adaptive frequency calculation method described above is then adopted to obtain the inversion frequencies and their corresponding frequency-domain Gaussian windows (Fig. 6).

First, we extract the time-frequency spectra of the pre-stack seismic data for the first inversion frequency $f_1$. The real and imaginary parts of these spectra are rearranged to construct a time-frequency spectral vector $\mathbf{d}_1$, and the time-frequency mapping matrix $\mathbf{\Psi}_1$ at this frequency is built. The model at this frequency is subsequently inverted (Fig. 7). The inverted model is then used as the prior model for the second inversion frequency $f_2$ to obtain the updated inverted result (Fig. 8). The above inversion procedure is repeated (intermediate inversion results are shown in Figs. 9 and 10) until all inversion frequencies are processed, yielding the final inverted model (Fig. 11). The inversion results indicate that the temporal resolution of the inverted model gradually improves with increasing inversion frequency, and the inverted model matches the true model better.

We further increase the noise level to an SNR of 10 dB. The time-frequency amplitude spectra of pre-stack seismic data at this SNR are displayed in Fig. 12, and the corresponding final inversion result is shown in Fig. 13. We use the reconstructed SNR to quantify the error of inversion results (Zhao et al., 2021):

$$\mathrm{SNR}_{\mathrm{dB}} = 10 \times \log_{10} \left( \frac{\sum_{n=1}^{N_t} \left[ \chi^{\mathrm{true}}(n) \right]^2}{\sum_{n=1}^{N_t} \left[ \chi^{\mathrm{inv}}(n) - \chi^{\mathrm{true}}(n) \right]^2} \right), \tag{30}$$

where $\chi^{\text{true}}$ and $\chi^{\text{inv}}$ represent true and inverted model, respectively.

The reconstructed SNRs of inversion results for seismic records with different signal-to-noise ratios are listed in Table 1. As the SNR decreases, the inversion accuracy degrades slightly. Nevertheless, the inverted results generally agree well with the true model, which demonstrates the effectiveness and robustness of the proposed inversion method.

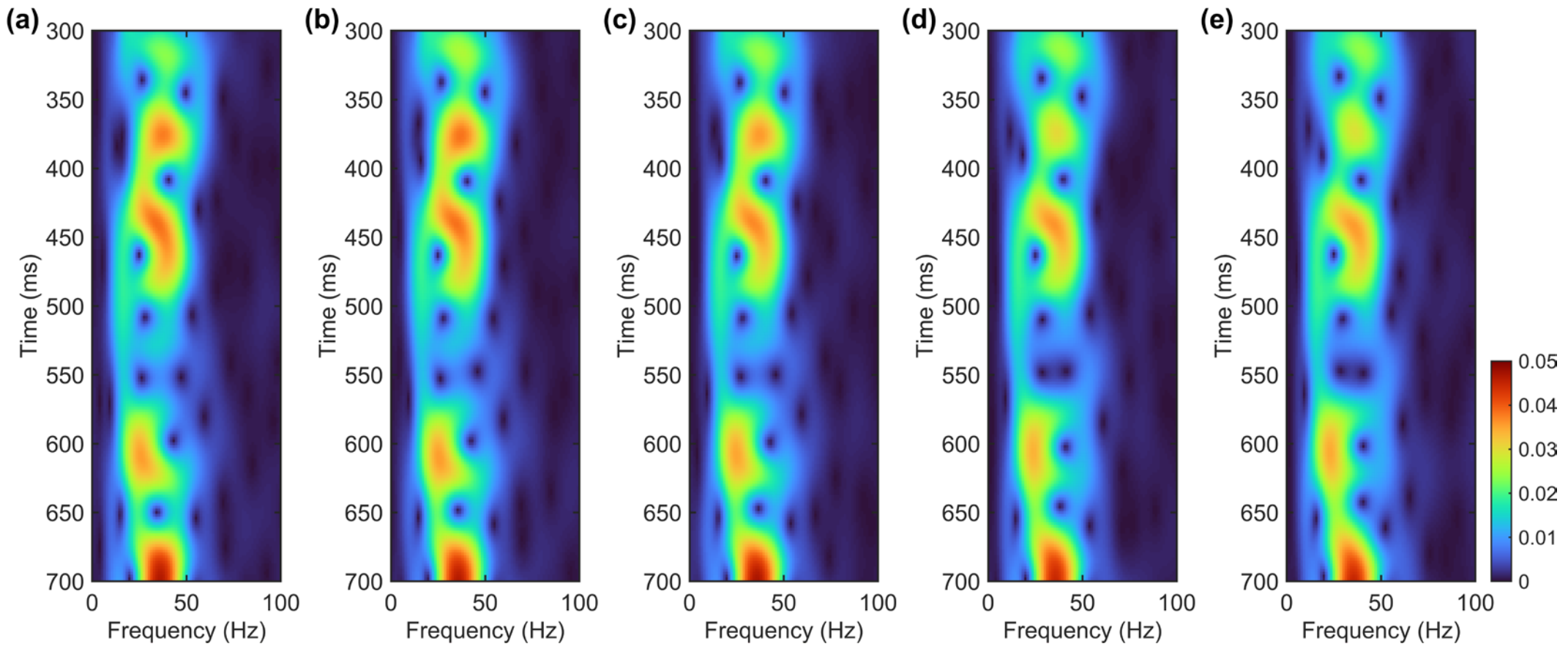

Fig. 5. Time-frequency amplitude spectra at different angles for an SNR of 20 dB.

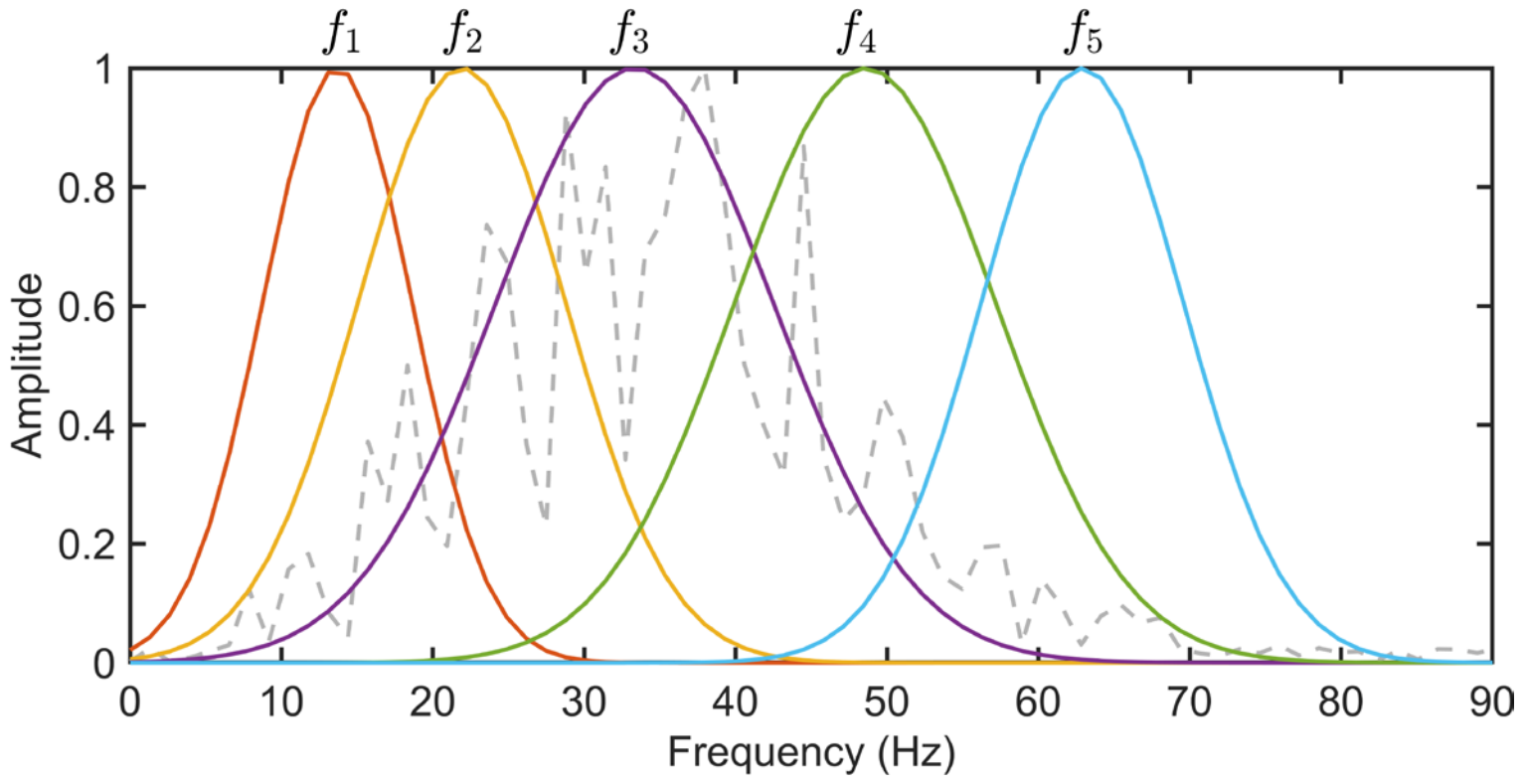

Fig. 6. Seismic spectrum and Gaussian window functions for different inversion frequencies.

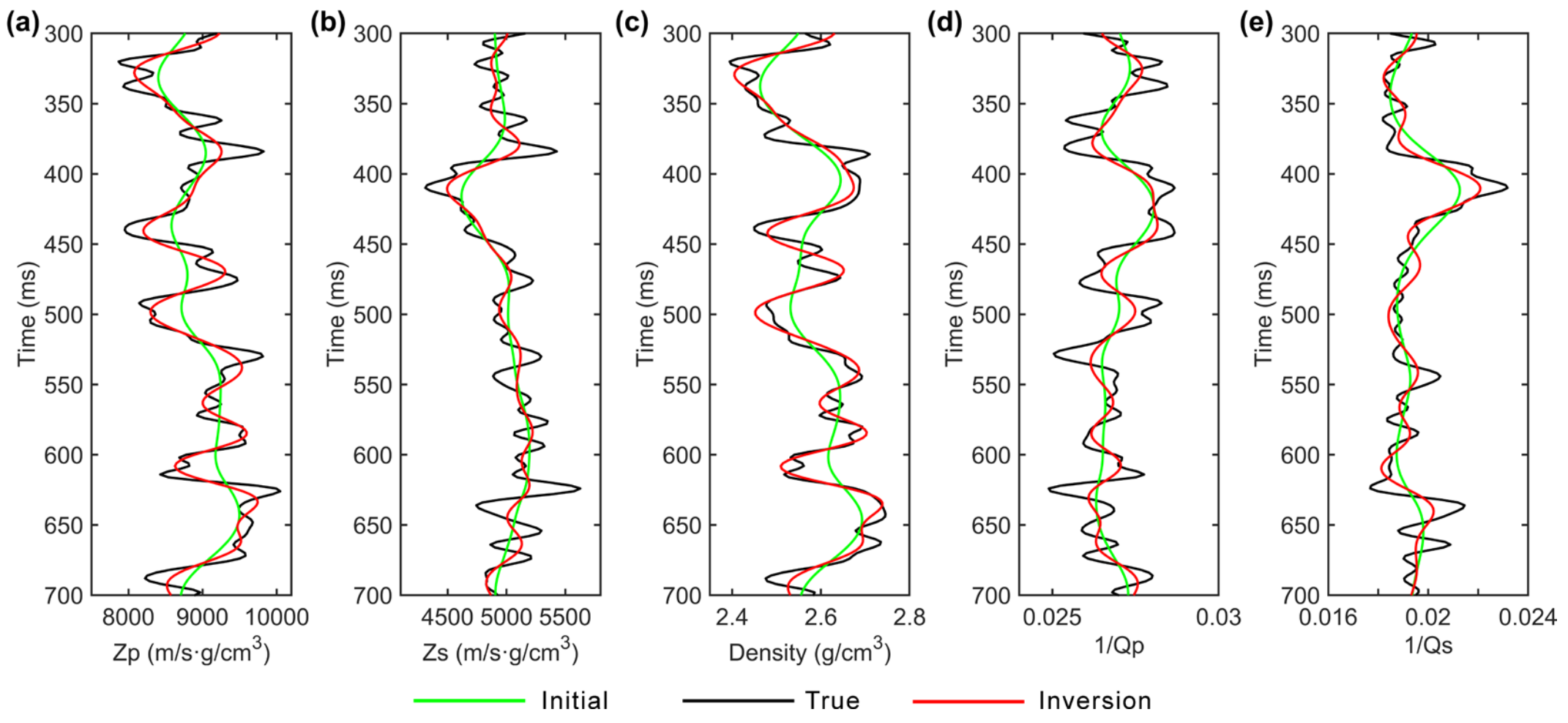

Fig. 7. Inversion results obtained by the inversion frequency vector $\mathbf{f} = [f_1]$ under an SNR of 20 dB. (a) P-wave impedance, (b) S-wave impedance, (c) density, (d) P-wave attenuation, and (e) S-wave attenuation.

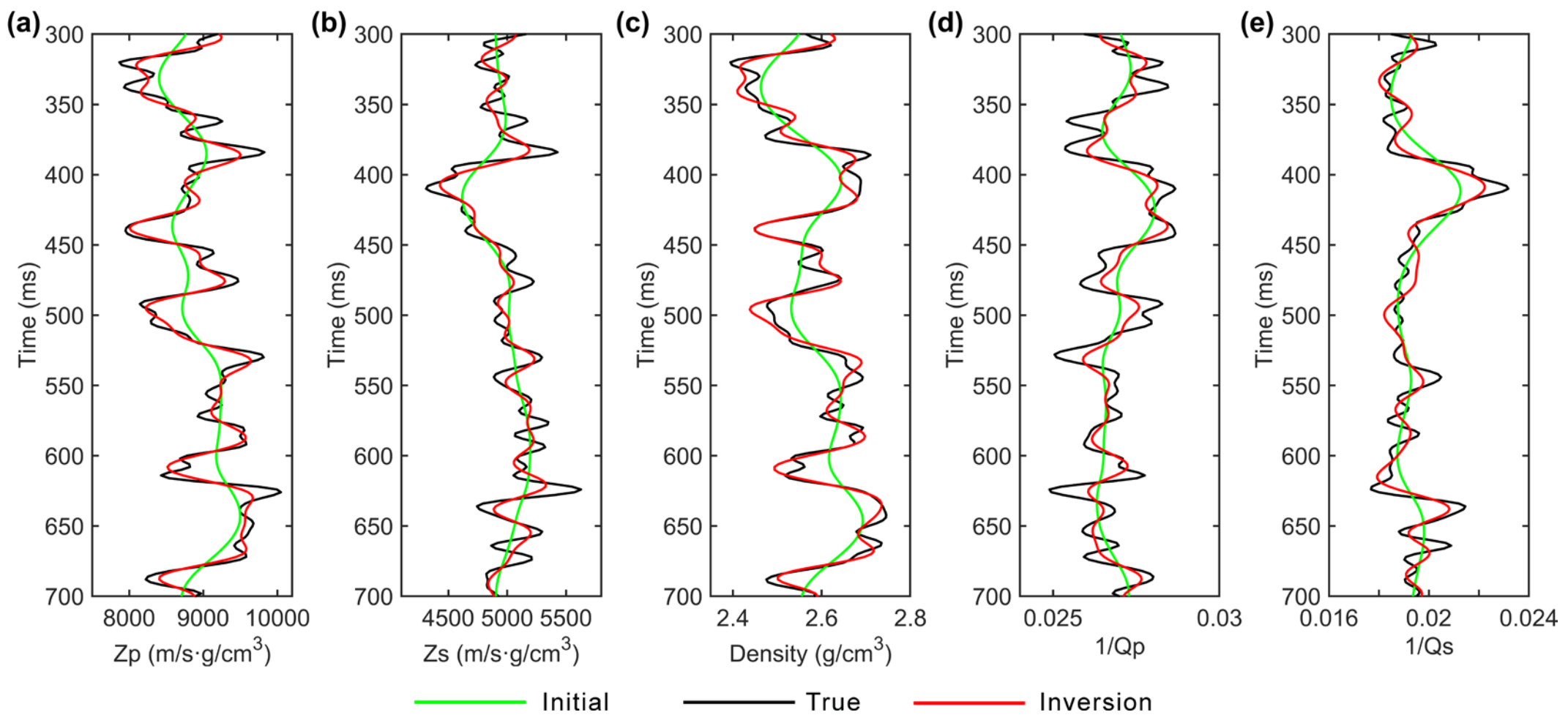

Fig. 8. Inversion results obtained by the inversion frequency vector $\mathbf{f} = [f_1, f_2]$ under an SNR of 20 dB. (a) P-wave impedance, (b) S-wave impedance, (c) density, (d) P-wave attenuation, and (e) S-wave attenuation.

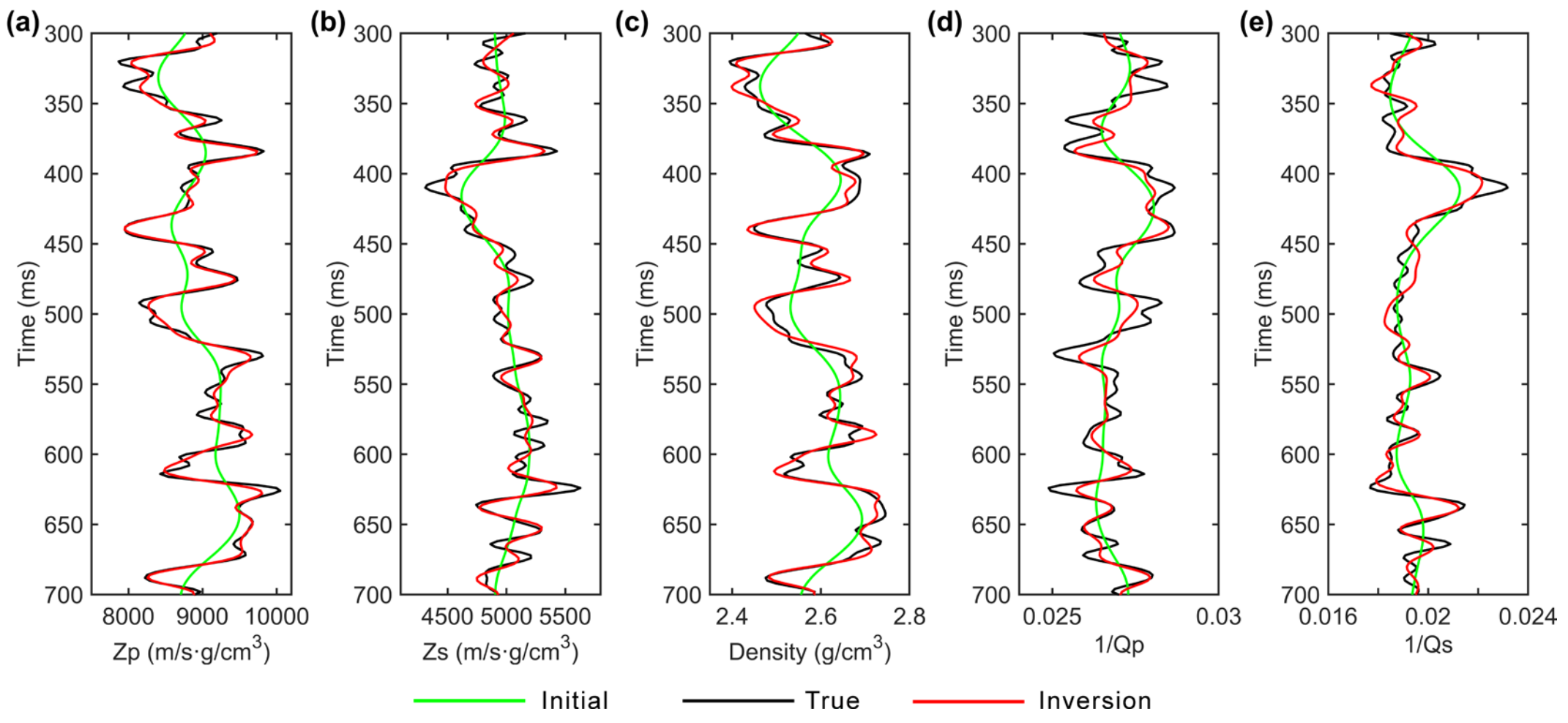

Fig. 9. Inversion results obtained by the inversion frequency vector $\mathbf{f} = [f_1, f_2, f_3]$ under an SNR of 20 dB. (a) P-wave impedance, (b) S-wave impedance, (c) density, (d) P-wave attenuation, and (e) S-wave attenuation.

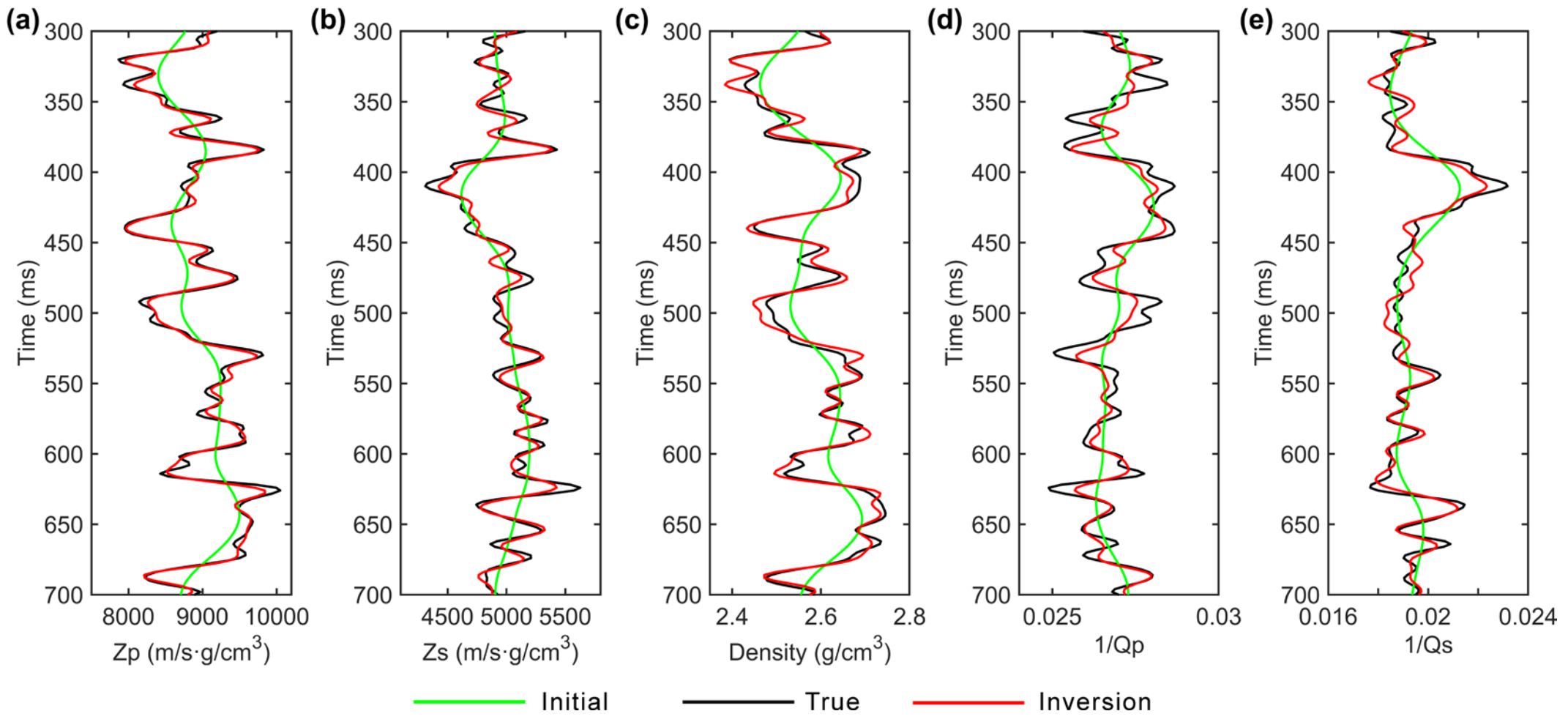

Fig. 10. Inversion results obtained by the inversion frequency vector $\mathbf{f} = [f_1, f_2, f_3, f_4]$ under an SNR of 20 dB. (a) P-wave impedance, (b) S-wave impedance, (c) density, (d) P-wave attenuation, and (e) S-wave attenuation.

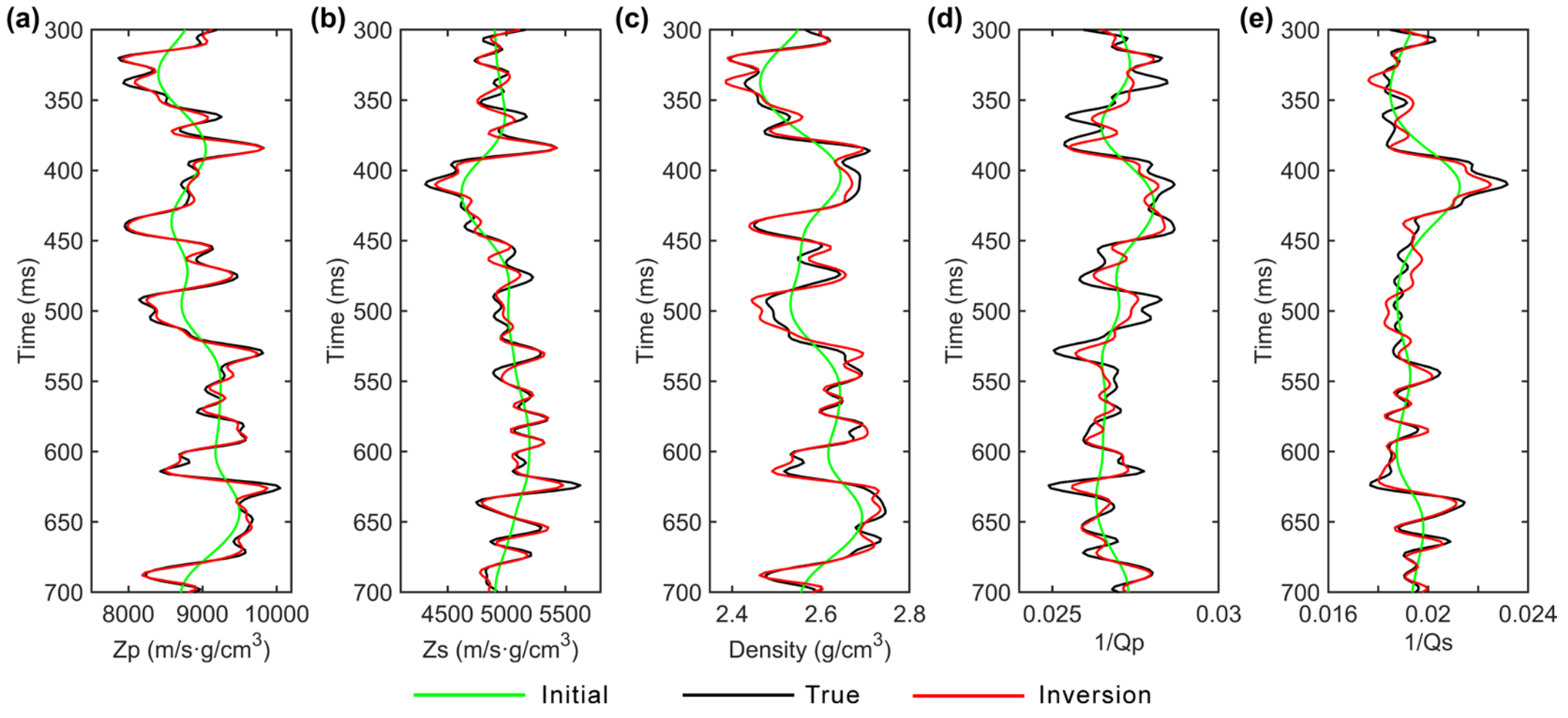

Fig. 11. Inversion results obtained by the inversion frequency vector $\mathbf{f} = [f_1, f_2, f_3, f_4, f_5]$ under an SNR of 20 dB. (a) P-wave impedance, (b) S-wave impedance, (c) density, (d) P-wave attenuation, and (e) S-wave attenuation.

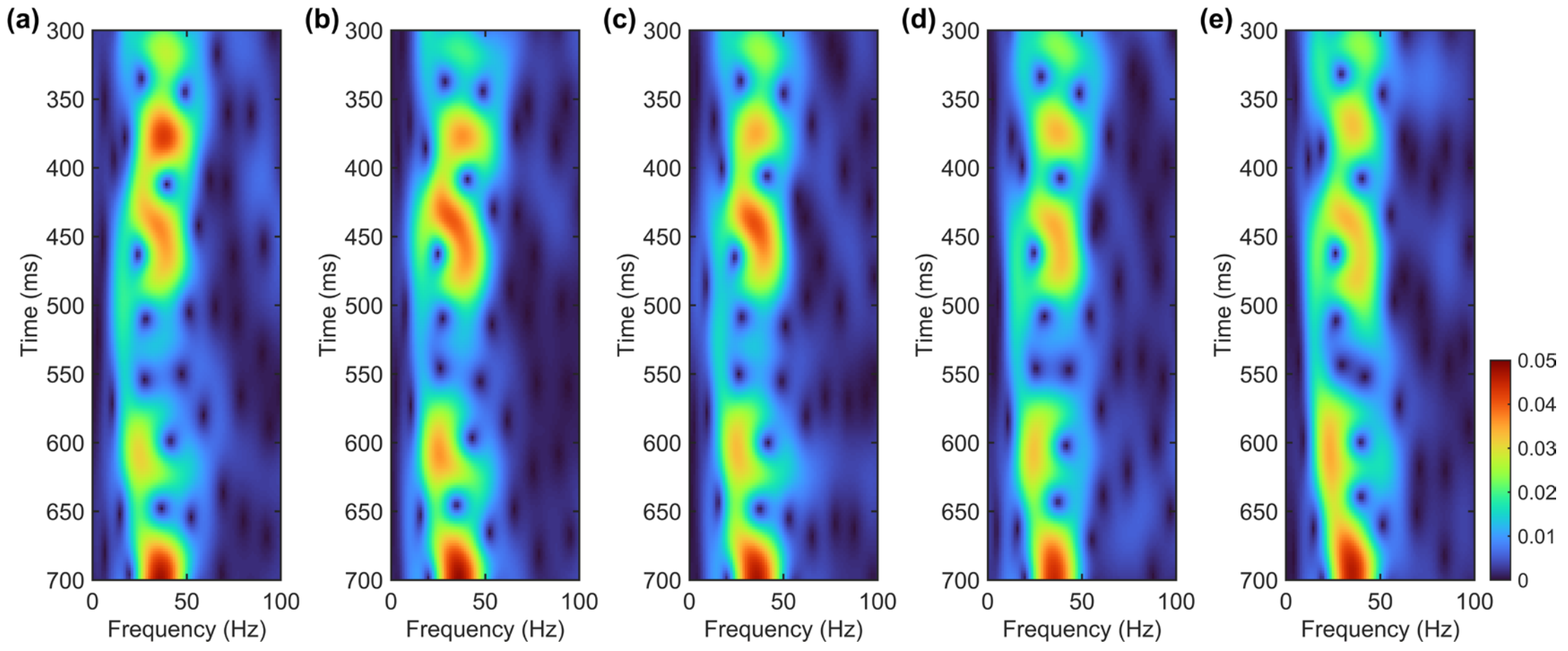

Fig. 12. Time-frequency amplitude spectra at different angles for an SNR of 10 dB.

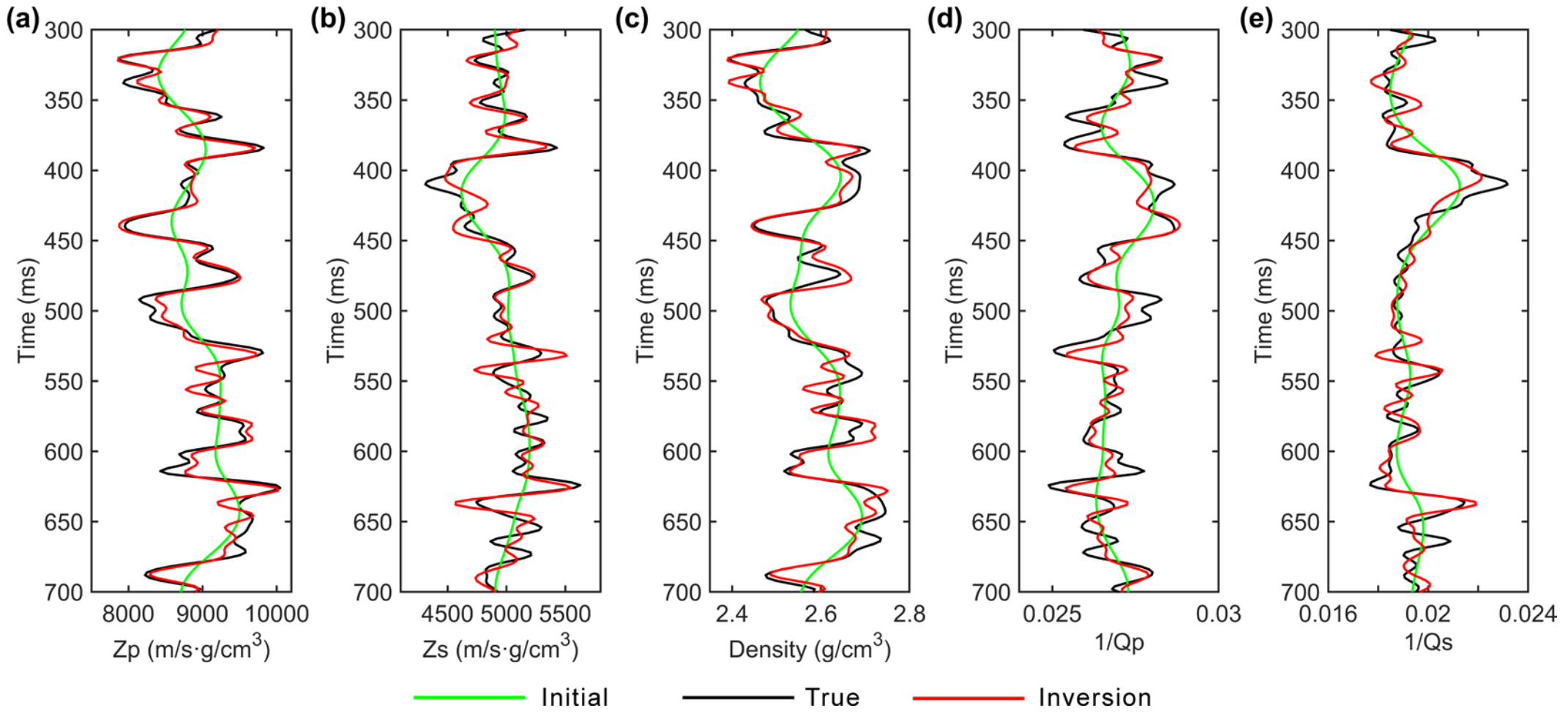

Fig. 13. Inversion results obtained by the inversion frequency vector

$\mathbf{f} = [f_1, f_2, f_3, f_4, f_5]$ under an SNR of 10 dB. (a) P-wave impedance, (b) S-wave impedance, (c) density, (d) P-wave attenuation, and (e) S-wave attenuation.

**Table 1. The reconstruction SNRs of the inverted model under different SNRs.**

| | $Z_\mathrm{p}$ | $Z_\mathrm{s}$ | $\rho$ | $Q_p^{-1}$ | $Q_s^{-1}$ |
|---|---|---|---|---|---|
| 20 dB | 41.07 | 38.12 | 41.93 | 36.74 | 34.48 |
| 10 dB | 36.20 | 33.31 | 39.19 | 35.88 | 31.55 |

# 4 Field Data Test

In this section, field seismic data are used to further test the effectiveness and applicability of the proposed method. Fig. 14 shows the pre-stack seismic sections at five incidence angles across Well 1 in an exploration area. Fig. 15 shows the seismic wavelets at different incidence angles at Well 1. Fig. 16 presents the well logs of Well 1. As can be seen from the gas saturation curve in Fig. 16g, there are three reservoir units in the interval of interest.

First, the critical porosity shown in Fig. 17c was obtained using the rock-physics inversion method described in Section 2.2, and the associated mineral component information is listed in Table 2. As shown in Figs. 17a-b, although the P- and S-wave velocities obtained from rock-physics modeling exhibit some local deviations from the logged P- and S-wave velocities, they generally agree well, confirming the accuracy of the inversion results. The logged P- and S-wave attenuations of Well 1 are then estimated through rock-physics modeling, as shown in Figs. 17d-g. These figures indicate that P-wave attenuation can effectively indicate reservoir locations: reservoir intervals exhibit strong attenuation, whereas non-reservoir intervals exhibit weak attenuation. In contrast, S-wave attenuation caused by rock-matrix heterogeneity is insensitive to fluids and cannot effectively indicate reservoir locations.

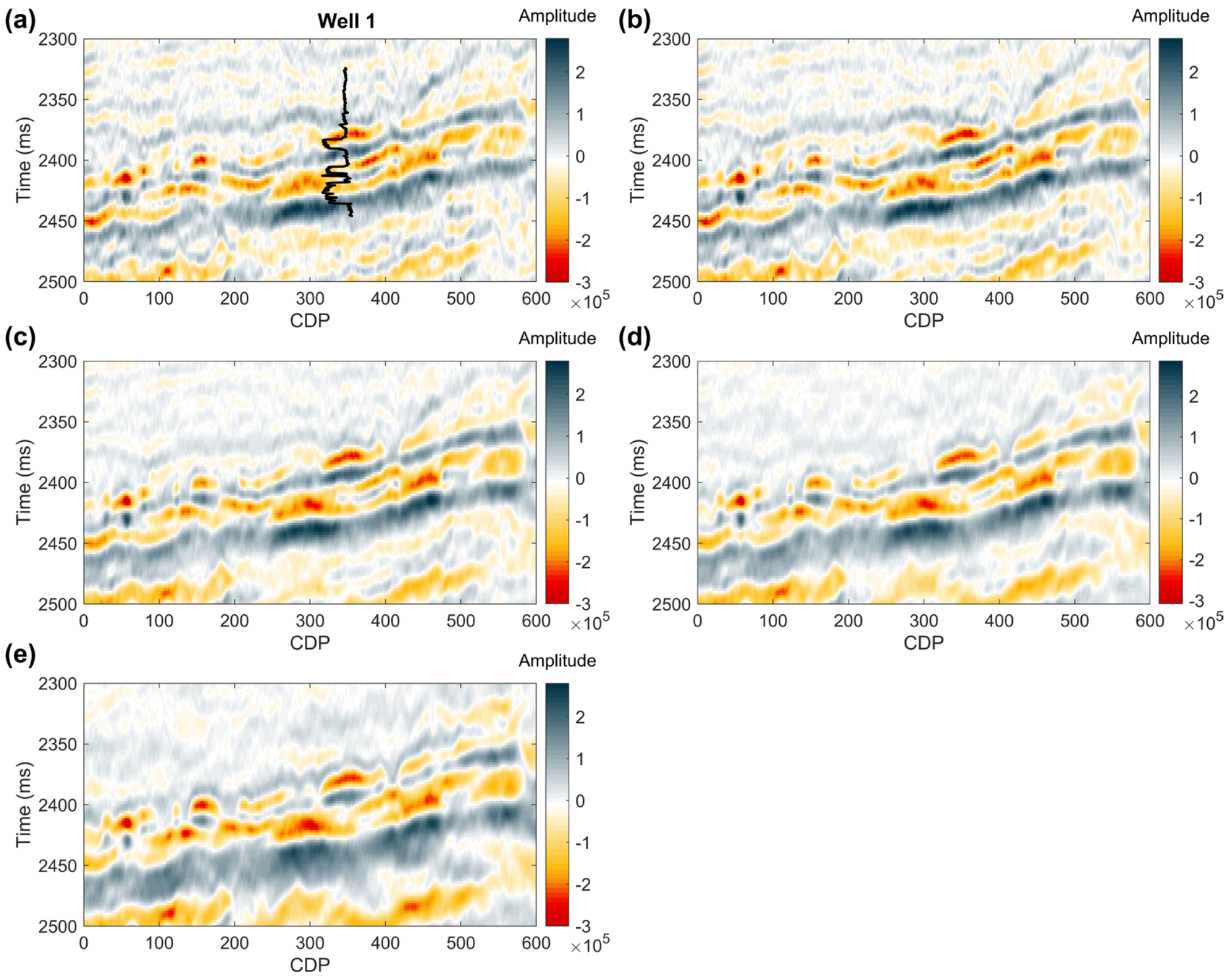

Fig. 14. Pre-stack seismic sections at different incidence angles across Well 1: (a) $\theta = 6^\circ$, (b) $\theta = 13^\circ$,(c) $\theta = 20^\circ$,(d) $\theta = 27^\circ$, and (e) $\theta = 34^\circ$. (The black curve in Fig. 14a denotes the P-wave impedance log.)

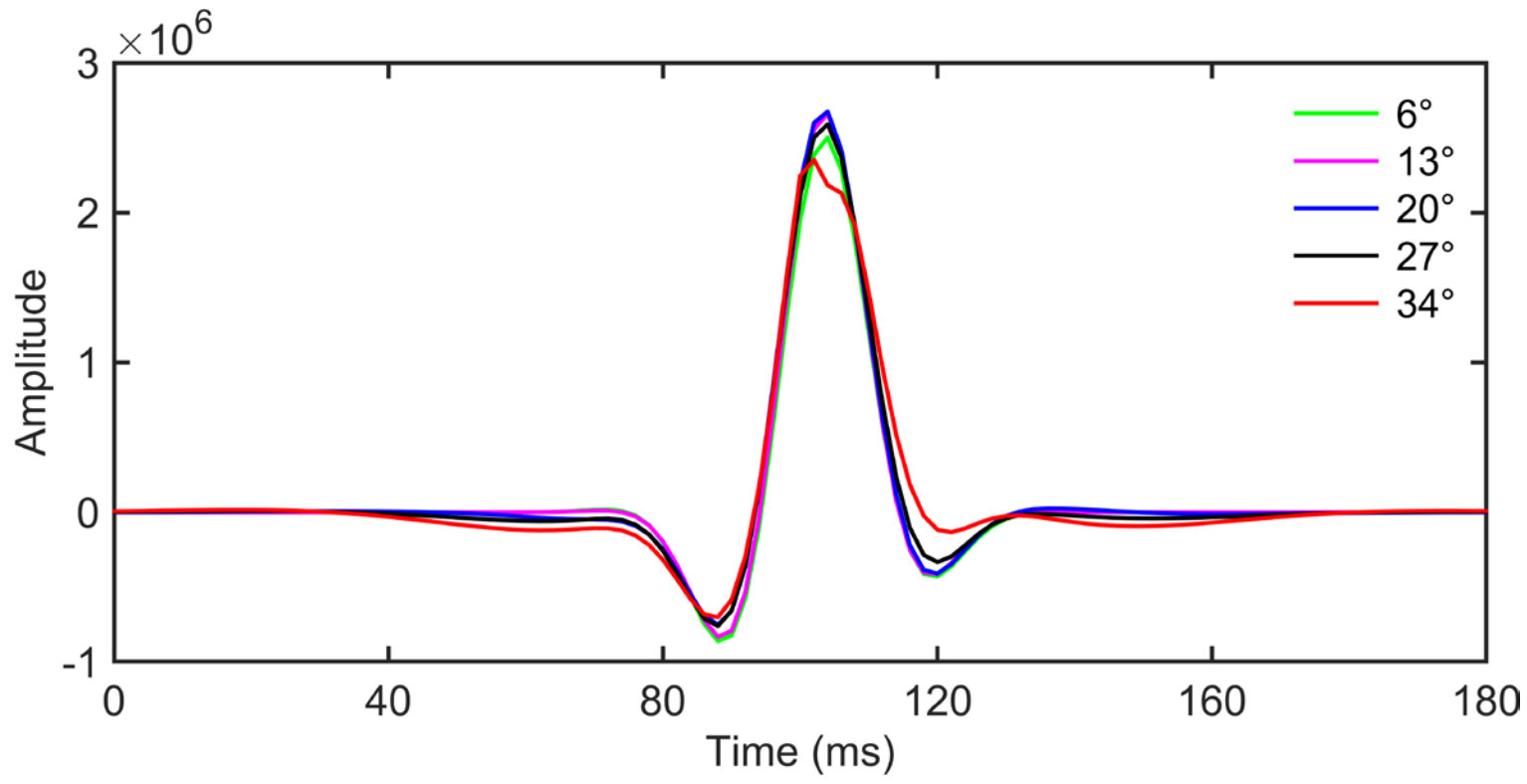

Fig. 15. Seismic wavelets at Well 1 for different incidence angles.

**Table 2. Summary of the bulk and shear moduli and density of all minerals used in the calculation.**

| Constituents | Bulk modulus (Gpa) | Shear modulus (Gpa) | Density (Kg/m$^3$) |
|---|---|---|---|
| Quartz | 38 | 44 | 2650 |
| Clay | 20.9 | 6.85 | 2550 |
| Water | 2.25 | 0 | 1.03 |
| Gas | 0.12 | 0 | 0.23 |

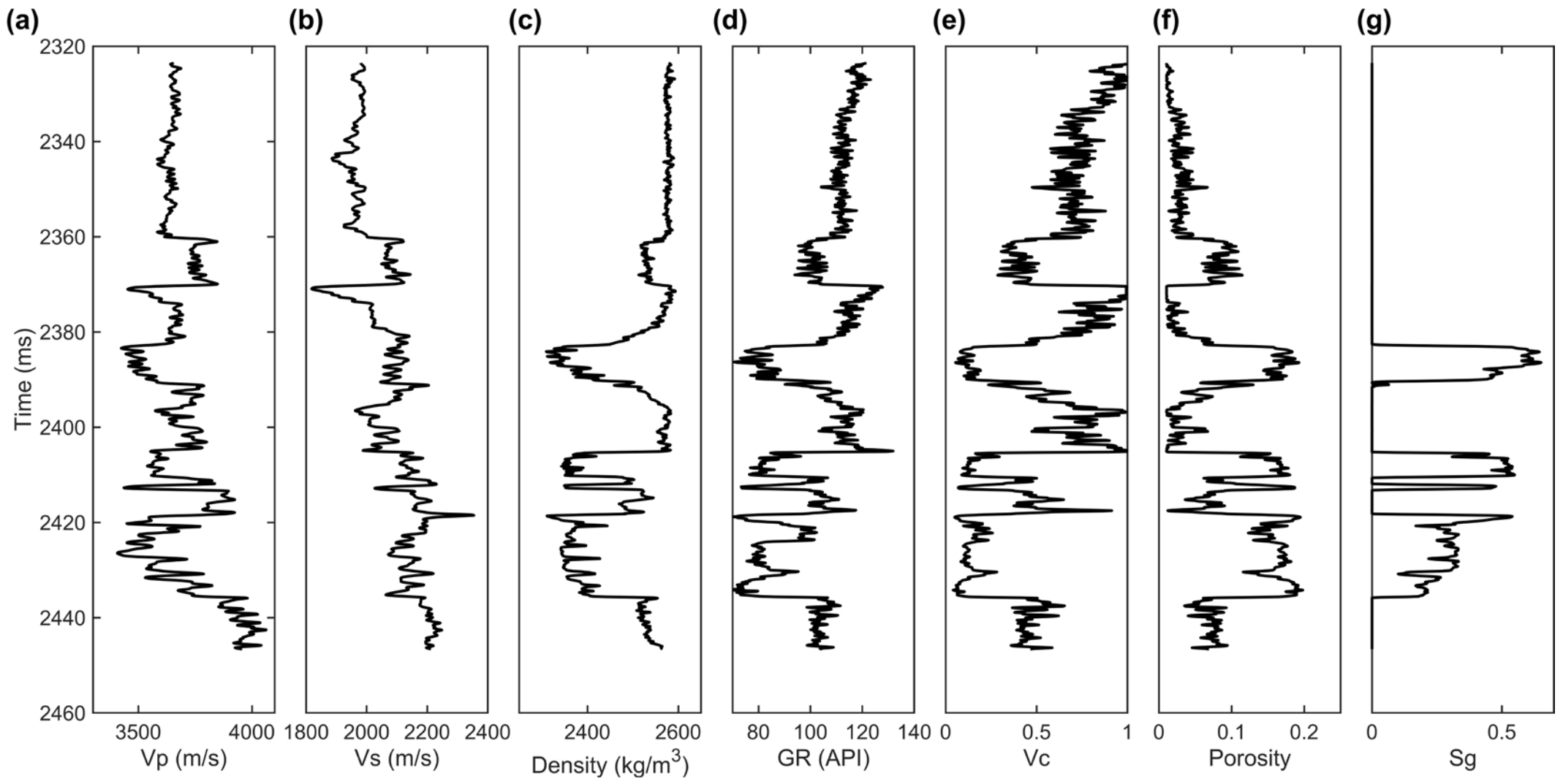

Fig. 16. Well logs of Well 1: (a) P-wave velocity, (b) S-wave velocity, (c) density, (d) Gamma, (e) shale content, (f) porosity, and (g) gas saturation.

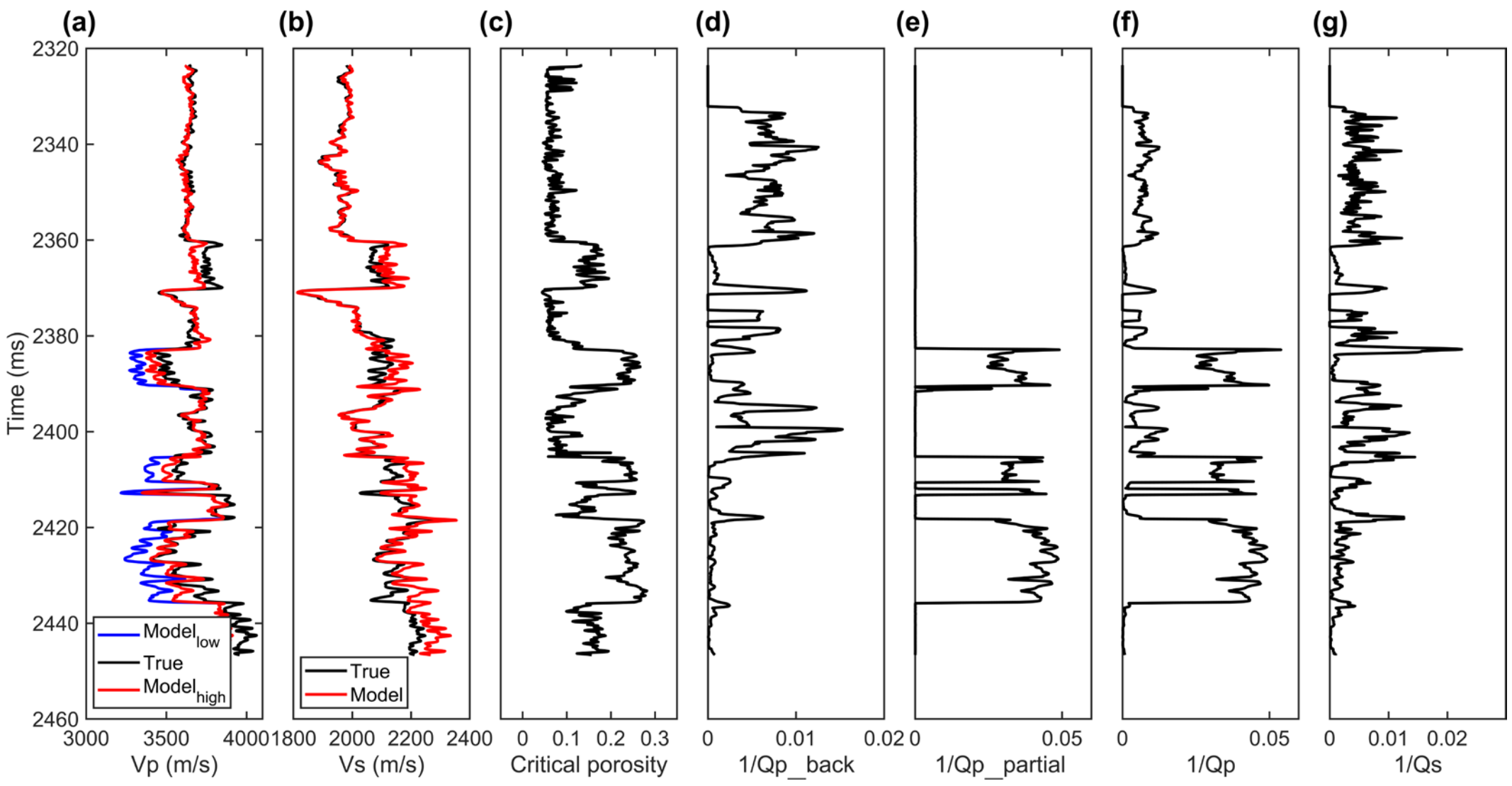

Fig. 17. Well-log inversion results for Well 1. (a) P-wave velocity, (b) S-wave velocity, (c) critical porosity, (d) background P-wave attenuation, (e) partial-saturation P-wave attenuation, (f) total P-wave attenuation, and (g) S-wave attenuation.

Second, the horizons, well logs, and the derived P- and S-wave attenuation logs are used to establish the prior models, as shown in Fig. 18. Since the cutoff frequency of the prior models is 10 Hz, the minimum inversion frequency was set to 15 Hz, and the inversion frequencies are then obtained using the adaptive inversion frequency calculation method. Fig. 19 shows the Gaussian window functions at different inversion frequencies and the normalized average amplitude spectrum of the seismic profile at an incidence angle of 5°.

Then, the time-frequency spectra of the pre-stack seismic data are computed. Taking the pre-stack seismic section at the incidence angle of 6° as an example, the time-frequency amplitude spectrum sections at different inversion frequencies are shown in Fig. 20. Finally, the final inversion sections are obtained using the time-frequency-domain Bayesian inversion method proposed in this paper, as shown in Fig.

21. As can be seen from the figures, P-wave impedance, density, and P-wave attenuation can effectively identify the three reservoir units. In the study area, reservoir intervals are characterized by low P-wave impedance, low density, and high P-wave attenuation. Among these attributes, P-wave impedance and P-wave attenuation provide the best reservoir prediction performance and, compared with density, can more clearly delineate the spatial distribution of the three reservoir units (as indicated by the red arrows in the figure). In contrast, reservoirs and non-reservoirs in the study area exhibit similar S-wave impedance and S-wave attenuation; therefore, S-wave impedance and S-wave attenuation cannot effectively identify reservoirs.

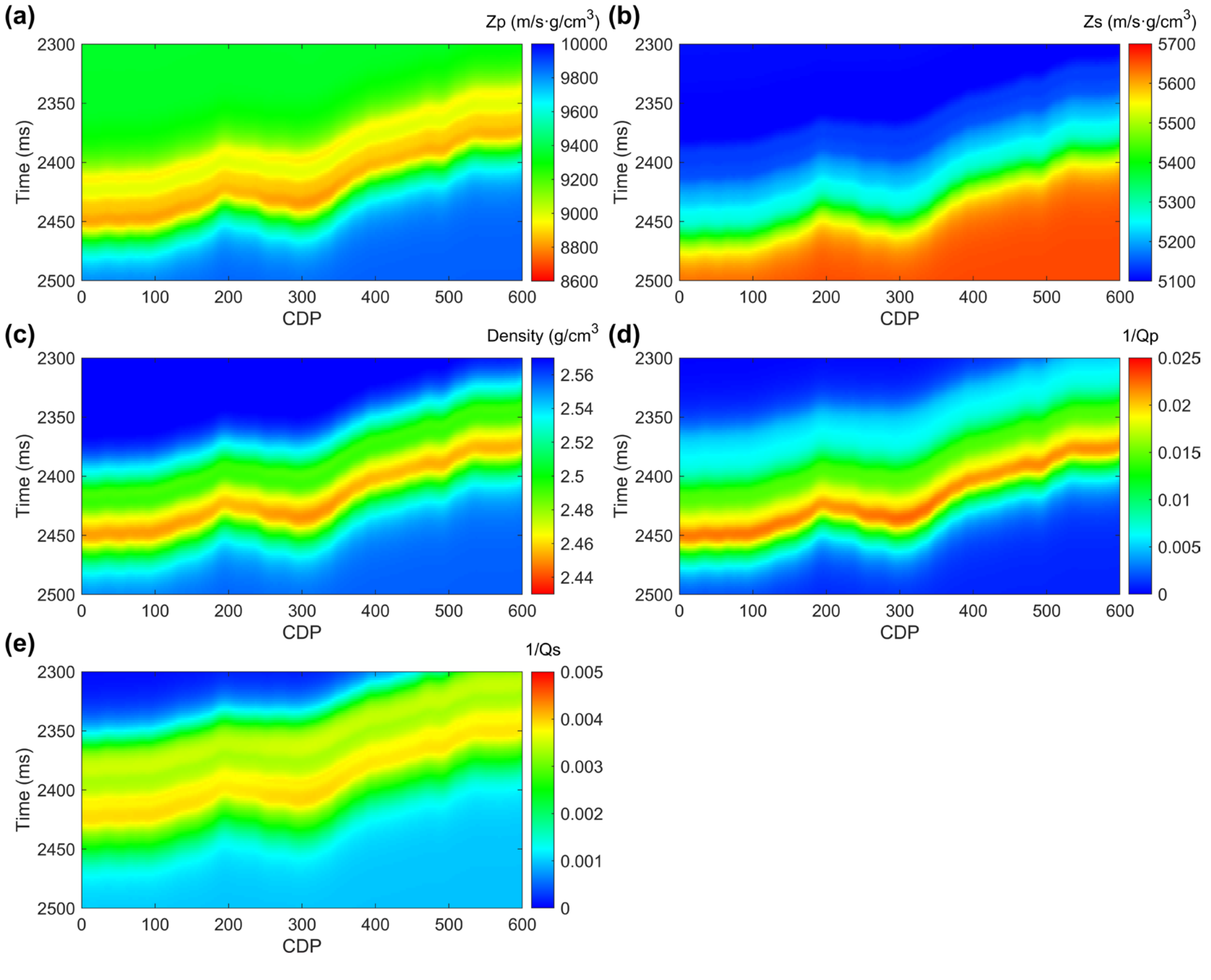

Fig. 18. Prior models. (a) P-wave impedance, (b) S-wave impedance, (c) density, (d) P-wave attenuation, and (e) S-wave attenuation.

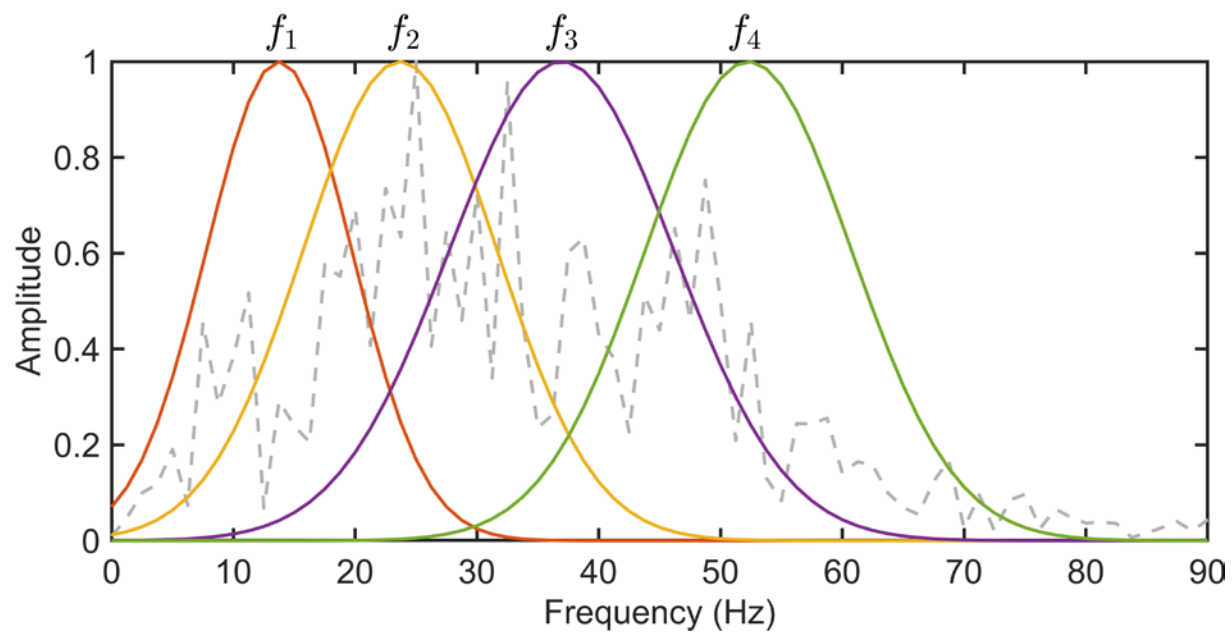

Fig. 19. Gaussian windows at different inversion frequencies. The gray curve denotes the normalized average amplitude spectrum of the seismic section at an incidence angle of 6°.

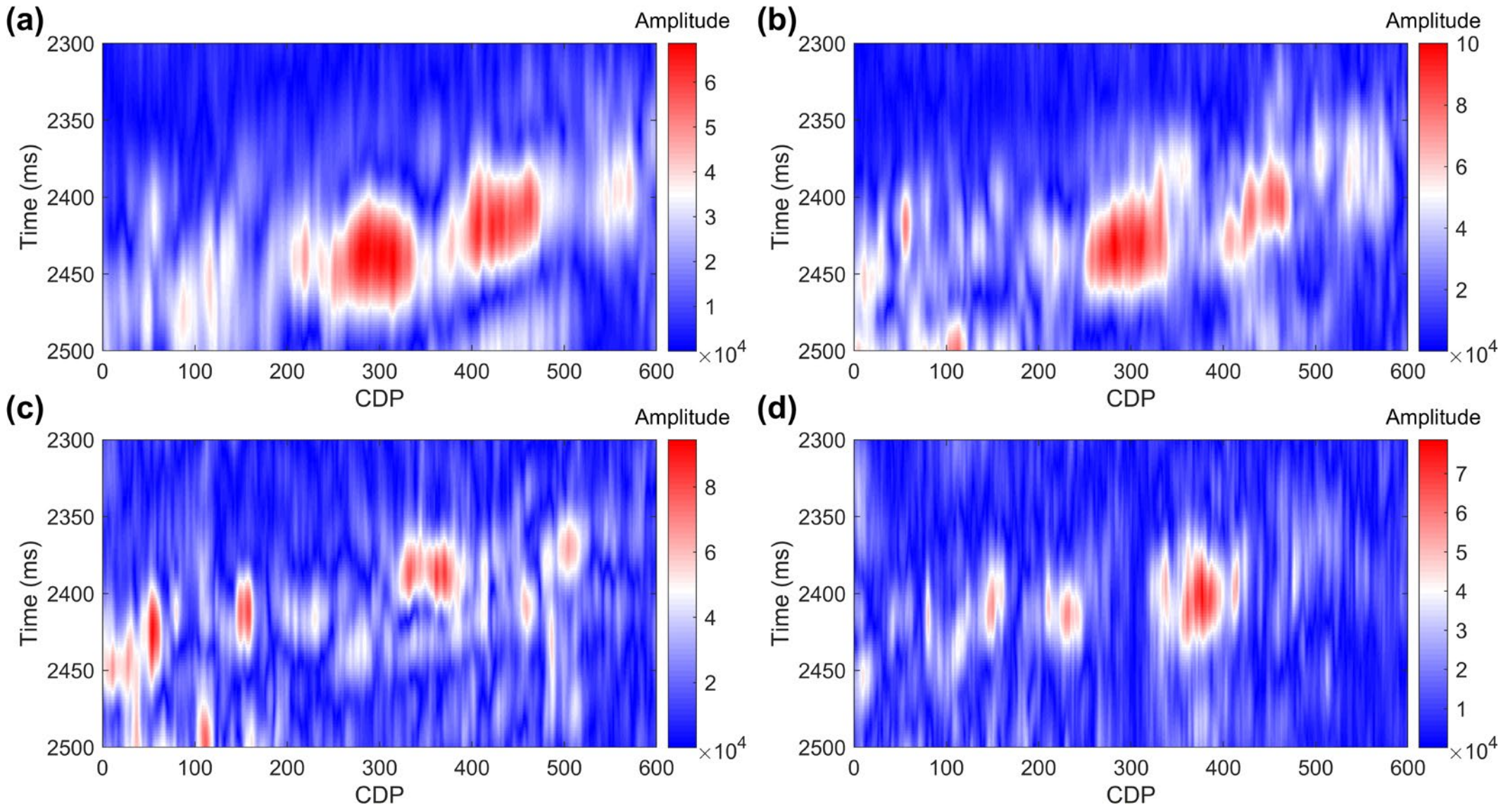

Fig. 20. Time-frequency amplitude spectra at different inversion frequencies for an incidence angle of 6°: (a) $f_1$, (b) $f_2$, (c) $f_3$, and $f_4$.

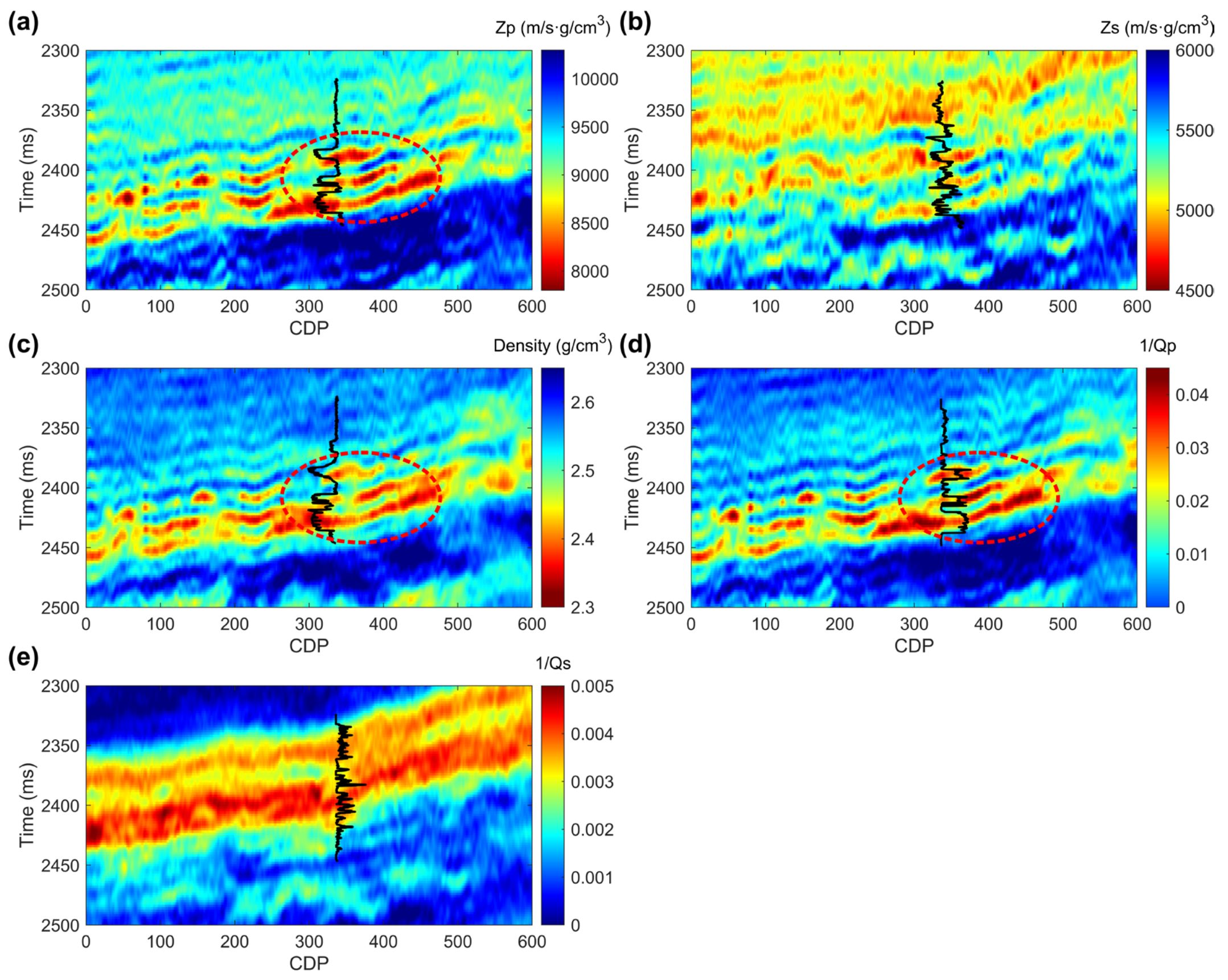

Fig. 21. Inversion result sections across Well 1. (a) P-wave impedance, (b) S-wave impedance, (c) density, (d) P-wave attenuation, and (e) S-wave attenuation.

Figs. 22-25 show the iterative inversion results at different inversion frequency vectors together with the actual well logs. As the inversion frequency increases, the resolution of the inverted P- and S-wave impedances, density, and P-wave attenuation gradually improves, and the inversion results show better agreement with the well-log values. In contrast, S-wave attenuation changes only slightly with increasing frequency. This is because S-wave attenuation is governed by rock-matrix heterogeneity and is insensitive to reservoirs, and the S-wave attenuation of the overall formation does not vary as markedly as P-wave attenuation. Fig. 25 presents the final inversion results; the good agreement between the inversion results and the well-log data confirms the effectiveness of the proposed method.

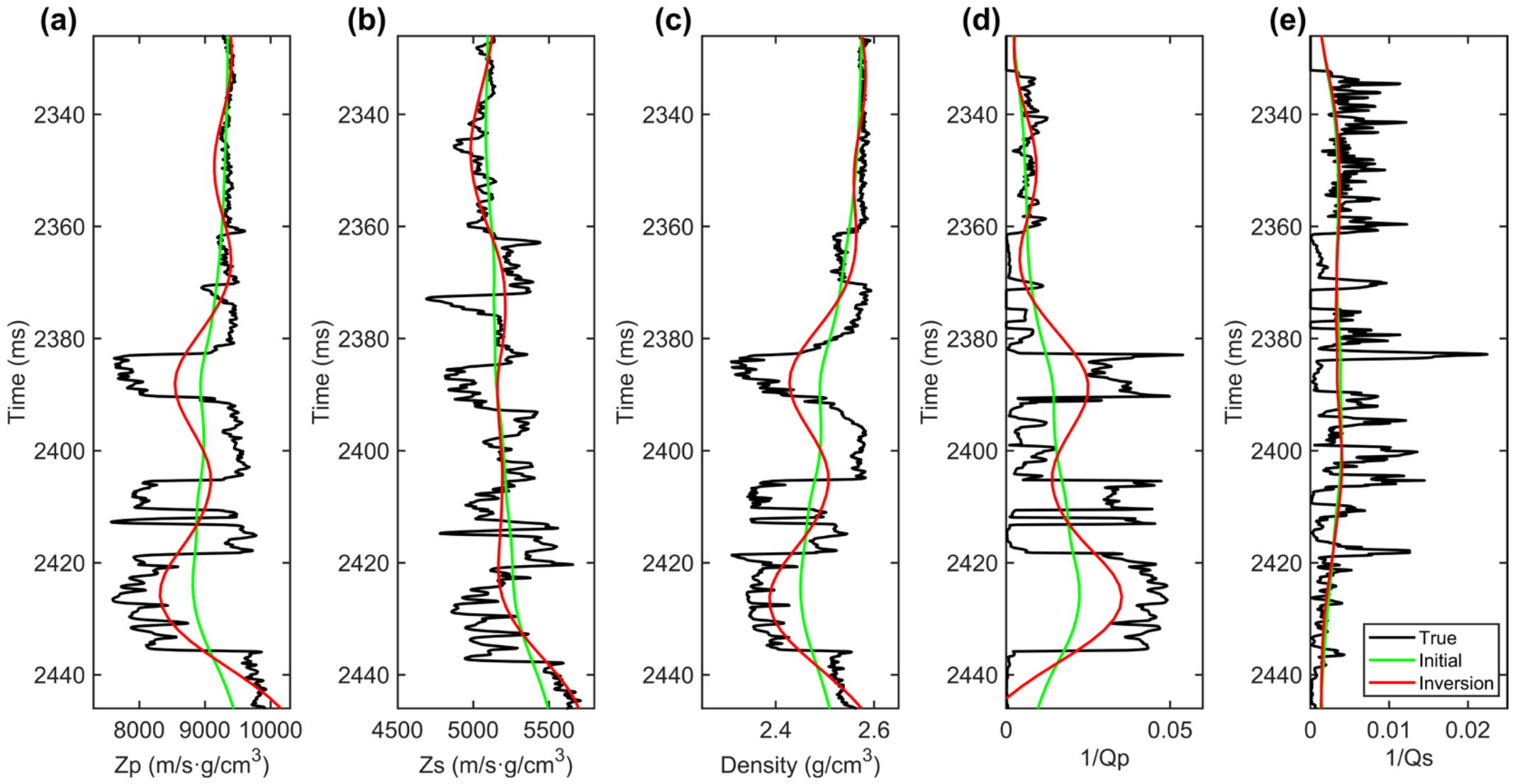

Fig. 22. Comparison between the inversion results near Well 1 and the well-log values for the inversion frequency vector $\mathbf{f} = [f_1]$. (a) P-wave impedance, (b) S-wave impedance, (c) density, (d) P-wave attenuation, and (e) S-wave attenuation.

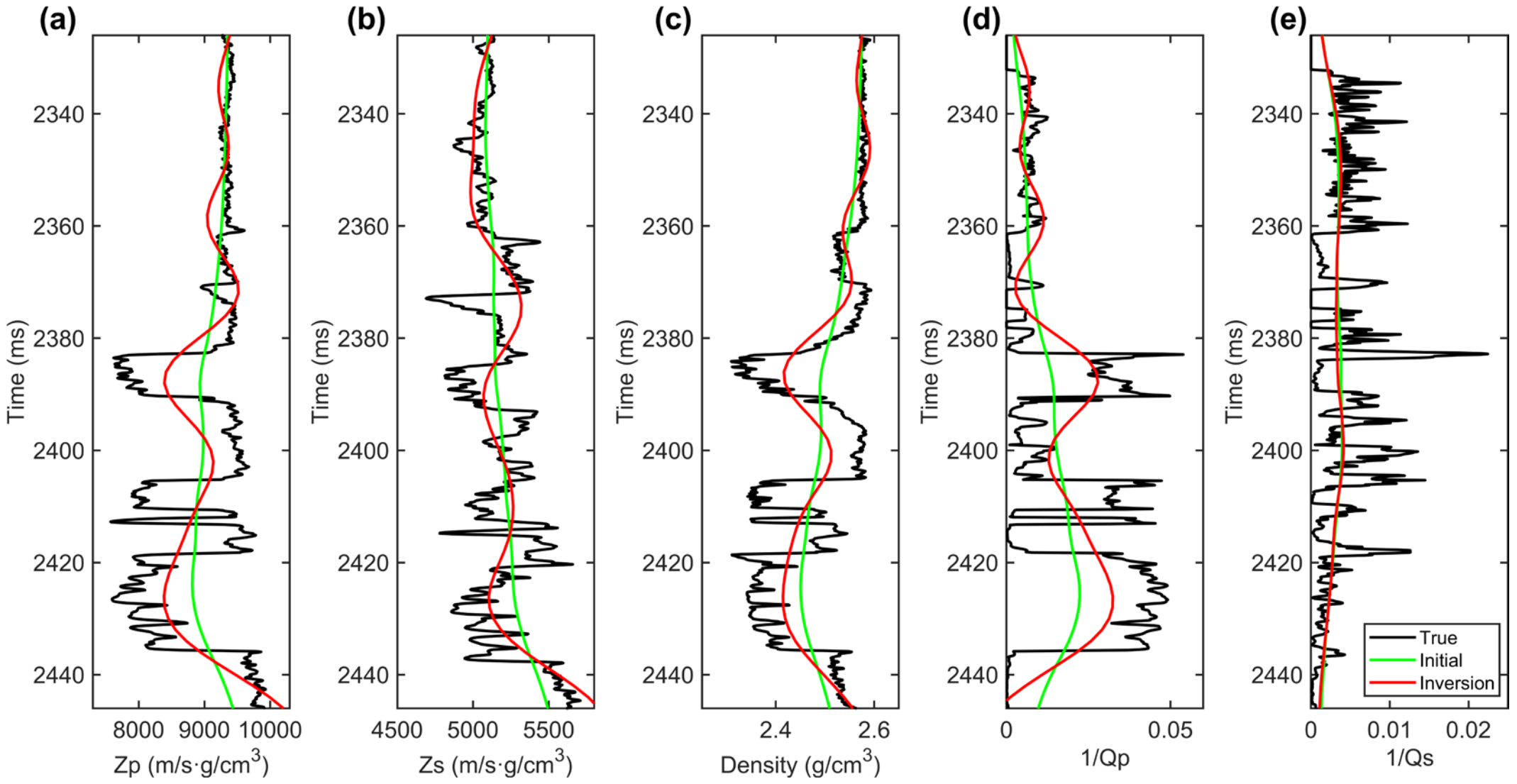

Fig. 23. Comparison between the inversion results near Well 1 and the well-log values for the inversion frequency vector $\mathbf{f} = [f_1, f_2]$. (a) P-wave impedance, (b) S-wave impedance, (c) density, (d) P-wave attenuation, and (e) S-wave attenuation.

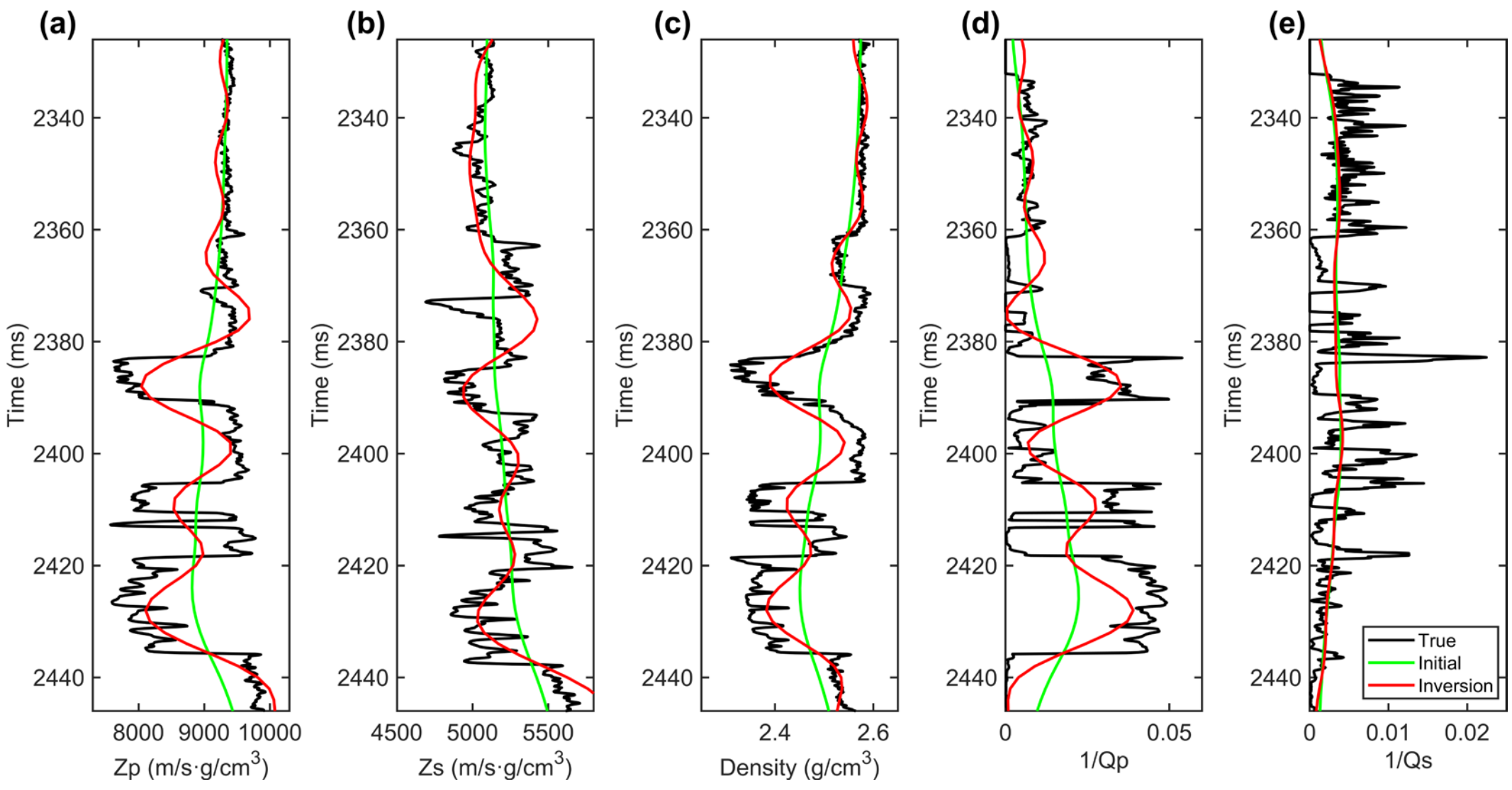

Fig. 24. Comparison between the inversion results near Well 1 and the well-log values for the inversion frequency vector $\mathbf{f} = [f_1, f_2, f_3]$. (a) P-wave impedance, (b) S-wave impedance, (c) density, (d) P-wave attenuation, and (e) S-wave attenuation.

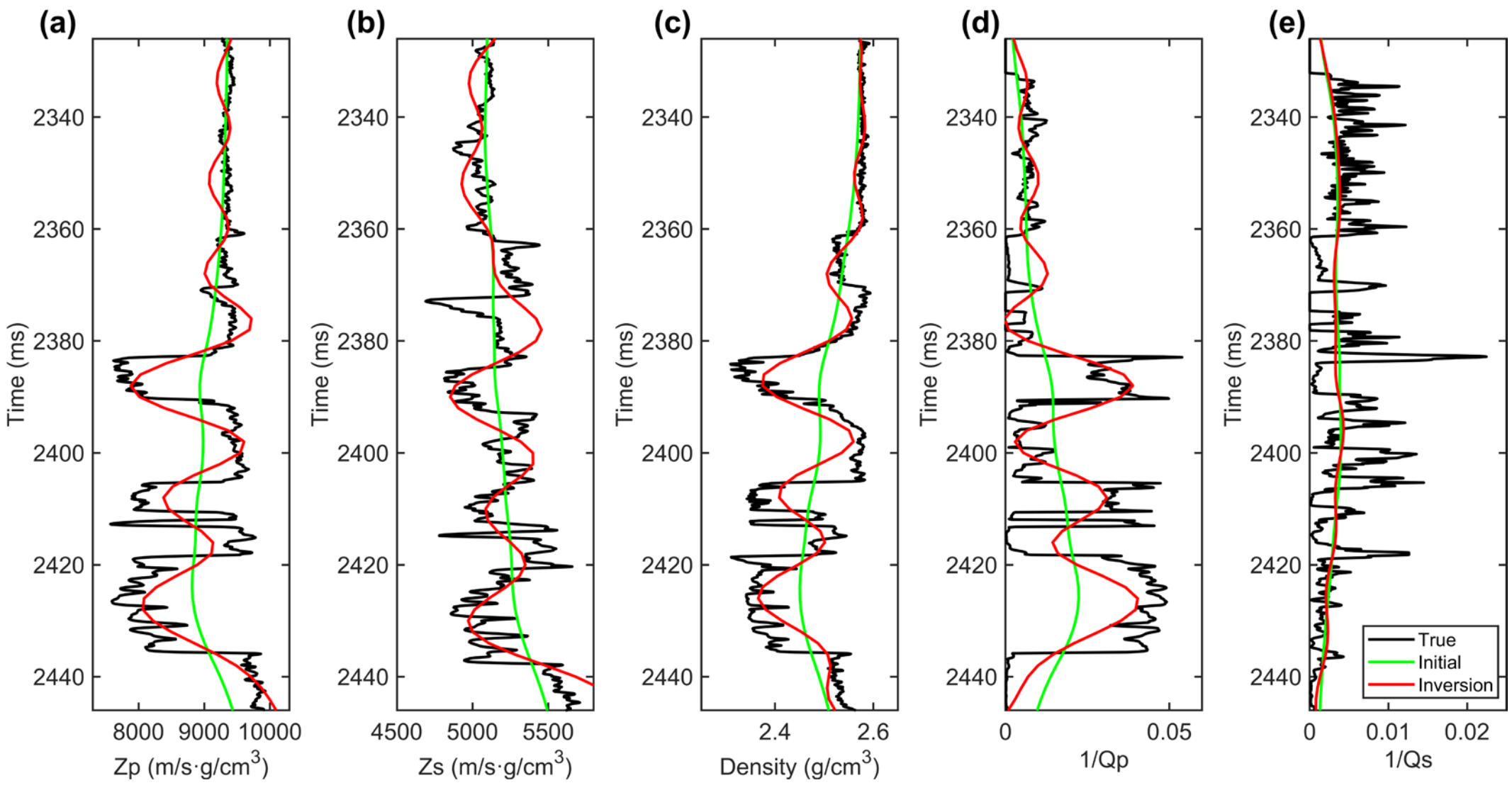

Fig. 25. Comparison between the inversion results near Well 1 and the well-log values for the inversion frequency vector $\mathbf{f} = [f_1, f_2, f_3, f_4]$. (a) P-wave impedance, (b) S-wave impedance, (c) density, (d) P-wave attenuation, and (e) S-wave attenuation.

# 5 Discussion

The KKR originates from the causality constraint of linear system response functions. Mathematically, it reflects the Hilbert transform relationship between the real and imaginary parts of the complex modulus arising from analyticity in the complex frequency domain (Carcione, 2022). Physically, the KKR indicates that seismic dispersion and attenuation in viscoelastic media are not independent processes; instead, they are jointly constrained by causality and constitute an inseparable coupled relationship (Mavko et al., 2020). Based on this principle, the present study employs the KKR to map medium attenuation into the dispersion structure of pre-stack reflection coefficients, thereby enabling frequency-dependent reflection modeling under causal constraints.

Several approximations underlying the derived equation should be acknowledged. The local KKR in Eq. (4) assumes that modulus dispersion varies slowly over the frequency range of interest, and the simplification from Eq. (11) to Eq. (12) further assumes that P- and S-wave attenuation are approximately constant near the reference angular frequency. These assumptions are reasonable for many rocks within the relatively narrow seismic frequency band, where attenuation often behaves approximately as constant. However, for materials with strong dispersion, such as rocks whose mesoscopic-flow characteristic frequency falls within the seismic band, attenuation may vary significantly with frequency. In such cases, the integral terms in Eq. (7) should be retained and Eq. (14) should be generalized accordingly.

Two additional assumptions are embedded in the linearization leading to Eq. (14): weak attenuation and weak contrasts. The weak attenuation assumption underlies the first-order expansion of the complex moduli that leads to Eq. (7), where products of attenuation perturbations are neglected. Appendix B shows that the imaginary part of the complex reflection coefficient is then small and contributes weakly to the total reflection coefficient. For stronger attenuation, higher-order terms or the exact complex reflection coefficient should be used. The weak contrast assumption underlies the linearization of the scattering function with respect to perturbations in stiffness and density, and limits Eq. (14) to interfaces with moderate impedance and attenuation contrasts. For strong attenuation and strong interface contrasts, a frequency-dependent nonlinear complex reflection coefficient incorporating both P- and S-wave attenuation should be considered. The target reservoirs in the examples have moderate contrasts, so the linearized equation is adequate; nevertheless, the weak contrast assumption should be checked against well logs before applying the method to stronger contrast areas.

The field data results show that P-wave attenuation anomalies identify gas reservoirs, whereas S-wave attenuation is insensitive to gas reservoirs. This suggests that P-wave attenuation is a more diagnostic attribute for gas reservoir prediction in the study area, while S-wave attenuation may contain limited fluid information. Nevertheless, S-wave attenuation is not necessarily uninformative in other settings. It can be sensitive to fractures, anisotropy, and matrix heterogeneity. Because the present formulation assumes isotropic linear viscoelastic media, anisotropic attenuation and frequency-dependent anisotropy are not considered. Extending the KKR-based reflection coefficient to anisotropic viscoelastic media, and combining it with rock-physics models that include fracture-induced dispersion and attenuation, would be a natural direction for future work.

Overall, the proposed method provides a physically grounded and numerically stable way to estimate P- and S-wave attenuation from pre-stack seismic data. Its main limitations are the weak attenuation and weakcontrast assumptions, the local-KKR and near-constant-attenuation approximations, and the dependence on well-log-based prior models. These limitations define clear targets for future improvements, including full-KKR or strong-contrast formulations, anisotropic viscoelastic extensions, and corresponding robust Bayesian inversion methods.

# 6 Conclusions

This study has developed a physically consistent frequency-dependent PP-wave reflection coefficient equation and a time-frequency domain Bayesian inversion method for estimating P- and S-wave attenuation from pre-stack seismic data. Starting from the Kramers-Kronig relations and scattering theory, the viscoelastic stiffness matrix is expressed in terms of seismic attenuation, and a compact frequency-dependent PP-wave reflection coefficient explicitly containing P- and S-wave attenuation terms is derived under the assumptions of weak attenuation and small elastic and attenuation contrasts across the interface. The derived equation satisfies the causality constraint and provides a suitable forward model for frequency-dependent AVO inversion of attenuation.

The inversion is implemented in the time-frequency domain using a recursive frequency strategy with half-power overlapping Gaussian windows. The result at the previous inversion frequency serves as the prior model for the next frequency, and low-frequency constraints are provided by well-log rock-physics modeling. Synthetic tests show that the temporal resolution and accuracy of the inverted P-wave impedance, S-wave impedance, density, and P- and S-wave attenuation improve as the inversion frequency increases. The method remains stable at a SNR of 10 dB, demonstrating its robustness under moderate noise contamination.

Application to field seismic data shows that P-wave attenuation anomalies effectively identify gas reservoirs, whereas S-wave attenuation is insensitive to gas reservoirs because it is mainly governed by rock-matrix heterogeneity. The inverted P-wave impedance and P-wave attenuation provide better reservoir delineation than density, and the inversion results agree well with well-log data. The proposed method therefore provides a new approach for extracting P- and S-wave attenuation information from pre-stack seismic data and for high-resolution gas reservoir prediction. The method is developed under several simplifying assumptions, including weak attenuation, weak scattering, and isotropic viscoelastic media. Its applicability to complex fractured reservoirs, strongly attenuating media propagation conditions remains to be further investigated. Future work could apply anisotropic viscoelastic theory to investigate attenuation related responses in frequency-dependent reflections under complex media conditions.

## Acknowledgments

This research was financially supported by the National Natural Foundation of China (No. 42374163) and the Key Program of Natural Science Foundation of Sichuan Province of China (No. 2026NSFSC1966).

## Appendix A: Derivation of the perturbation of the complex p-wave and shear modulus

The perturbation of the real part of the complex P-wave modulus $\tilde{M}(\omega)$ is given by

$$\begin{aligned}\Delta M_R(\omega) &= \left[M_R(\omega_0)+\Delta M_R(\omega_0)\right]\left\{1+\frac{2}{\pi}\int_{\ln\omega_0}^{\ln\omega}\left[Q_P^{-1}(\omega')+\Delta Q_P^{-1}(\omega')\right]d(\ln\omega')\right\} \\ &\quad -M_R(\omega_0)\left[1+\frac{2}{\pi}\int_{\ln\omega_0}^{\ln\omega}Q_P^{-1}(\omega')d(\ln\omega')\right] \\ &= \Delta M_R(\omega_0)+\frac{2}{\pi}M_R(\omega_0)\int_{\ln\omega_0}^{\ln\omega}\Delta Q_P^{-1}(\omega')d(\ln\omega') \\ &\quad +\frac{2}{\pi}\Delta M_R(\omega_0)\int_{\ln\omega_0}^{\ln\omega}\left[Q_P^{-1}(\omega')+\Delta Q_P^{-1}(\omega')\right]d(\ln\omega')\end{aligned} \quad . \quad \text{(A1)}$$

Under the assumptions of weak contrasts in elastic properties across the interface and weak attenuation, Eq. B1 can be approximately simplified as:

$$\Delta M_R(\omega)\approx\Delta M_R(\omega_0)+\frac{2}{\pi}M_R(\omega_0)\int_{\ln\omega_0}^{\ln\omega}\Delta Q_P^{-1}(\omega')d(\ln\omega'). \quad \text{(A2)}$$

The perturbation of the imaginary part of the complex P-wave modulus $\tilde{M}(\omega)$ is expressed as:

$$\Delta M_I(\omega)=\left[M_R(\omega_0)+\Delta M_R(\omega_0)\right]\left\{1+\frac{2}{\pi}\int_{\ln\omega_0}^{\ln\omega}\left[Q_P^{-1}(\omega')+\Delta Q_P^{-1}(\omega')\right]d(\ln\omega')\right\}\left[Q_P^{-1}(\omega)+\Delta Q_P^{-1}(\omega)\right]$$
$$-M_R(\omega_0)\left[1+\frac{2}{\pi}\int_{\ln\omega_0}^{\ln\omega}Q_P^{-1}(\omega')d(\ln\omega')\right]Q_P^{-1}(\omega)$$
$$=M_R(\omega_0)\Delta Q_P^{-1}(\omega)+\Delta M_R(\omega_0)Q_P^{-1}(\omega)+\Delta M_R(\omega_0)\Delta Q_P^{-1}(\omega)$$
$$+\frac{2}{\pi}\left[M_R(\omega_0)\Delta Q_P^{-1}(\omega)+\Delta M_R(\omega_0)Q_P^{-1}(\omega)+\Delta M_R(\omega_0)\Delta Q_P^{-1}(\omega)\right]\int_{\ln\omega_0}^{\ln\omega}Q_P^{-1}(\omega')d(\ln\omega')$$
$$+\frac{2}{\pi}\left[M_R(\omega_0)Q_P^{-1}(\omega)+M_R(\omega_0)\Delta Q_P^{-1}(\omega)+\Delta M_R(\omega_0)Q_P^{-1}(\omega)+\Delta M_R(\omega_0)\Delta Q_P^{-1}(\omega)\right]\int_{\ln\omega_0}^{\ln\omega}\Delta Q_P^{-1}(\omega')d(\ln\omega') .$$

(A3)

Eq. B3 can be further approximated as:

$$\Delta M_I(\omega)\approx M_R(\omega_0)\Delta Q_P^{-1}(\omega)$$
$$+\frac{2}{\pi}M_R(\omega_0)\left[\Delta Q_P^{-1}(\omega)\int_{\ln\omega_0}^{\ln\omega}Q_P^{-1}(\omega')d(\ln\omega')+Q_P^{-1}(\omega)\int_{\ln\omega_0}^{\ln\omega}\Delta Q_P^{-1}(\omega')d(\ln\omega')\right] .$$

(A4)

Finally, the perturbation of the complex P-wave modulus $\tilde{M}(\omega)$ can be written as:

$$\Delta\tilde{M}(\omega)\approx\Delta M_R(\omega_0)+M_R(\omega_0)\frac{2}{\pi}\int_{\ln\omega_0}^{\ln\omega}\Delta Q_P^{-1}(\omega')d(\ln\omega')$$
$$+i\left\{M_R(\omega_0)\Delta Q_P^{-1}(\omega)+\frac{2}{\pi}M_R(\omega_0)\left[\Delta Q_P^{-1}(\omega)\int_{\ln\omega_0}^{\ln\omega}Q_P^{-1}(\omega')d(\ln\omega')+Q_P^{-1}(\omega)\int_{\ln\omega_0}^{\ln\omega}\Delta Q_P^{-1}(\omega')d(\ln\omega')\right]\right\}$$

.(A5)

Following a similar derivation procedure, the perturbation of the complex shear modulus $\tilde{\mu}(\omega)$ can be obtained as:

$$\Delta\tilde{\mu}(\omega)\approx\Delta\mu_R(\omega_0)+\mu(\omega_0)\frac{2}{\pi}\int_{\ln\omega_0}^{\ln\omega}\Delta Q_S^{-1}(\omega')d(\ln\omega')$$
$$+i\left\{\mu_R(\omega_0)\Delta Q_S^{-1}(\omega)+\frac{2}{\pi}\mu_R(\omega_0)\left[\Delta Q_S^{-1}(\omega)\int_{\ln\omega_0}^{\ln\omega}Q_S^{-1}(\omega')d(\ln\omega')+Q_S^{-1}(\omega)\int_{\ln\omega_0}^{\ln\omega}\Delta Q_S^{-1}(\omega')d(\ln\omega')\right]\right\}$$

.(A6)

## Appendix B: Analysis of the complex reflection coefficient in Eq. 10

Assume a reflection interface. The upper layer is an isotropic poroelastic formation, whose poroelastic characteristics are governed by Chapman's microscopic poroelastic model (Chapman et al., 2002, 2006). The basic rock-physics parameters of this lower formation are listed in Table B1. The lower layer is an isotropic, elastic formation, with its elastic parameters listed in Table B2.

By varying the crack density of the poroelastic formation while keeping the other parameters unchanged, we vary the P- and S-wave attenuation and analyze the reflection-amplitude characteristics of Eq. 10 under different P- and S-wave attenuations. As shown in Figs. B1–B4, under weak attenuation, the absolute value of the imaginary part of Eq. 10 increases with increasing P- and S-wave attenuation. However, this absolute value is small and close to zero, markedly smaller than that of the real part, and thus contributes weakly to the overall reflection coefficient. Therefore, in seismic inversion, the imaginary part of Eq. 10 can be neglected under the weak attenuation assumption.

**Table B1. The rock-physics parameters of the upper layer**

| Physical parameter | Value |
|---|---|
| Grain bulk modulus (GPa) | 13 |
| Grain shear modulus (GPa) | 8 |
| Gas bulk modulus (GPa) | 0.012 |
| Water bulk modulus (GPa) | 2.25 |
| Grain density (g/cm$^3$) | 2.35 |
| Gas density (g/cm$^3$) | 0.078 |
| Water density (g/cm$^3$) | 1.04 |
| Porosity | 0.15 |
| Crack aspect ratio | 0.01 |
| Gas saturation | 0.1 |
| Timescale parameter | 0.006 |

**Table B2. The physical parameters of the lower layer**

| Physical parameter | Value |
|---|---|
| P-wave velocity (m/s) | 3000 |
| S-wave velocity (m/s) | 1700 |
| Density (g/cm$^3$) | 2.2 |

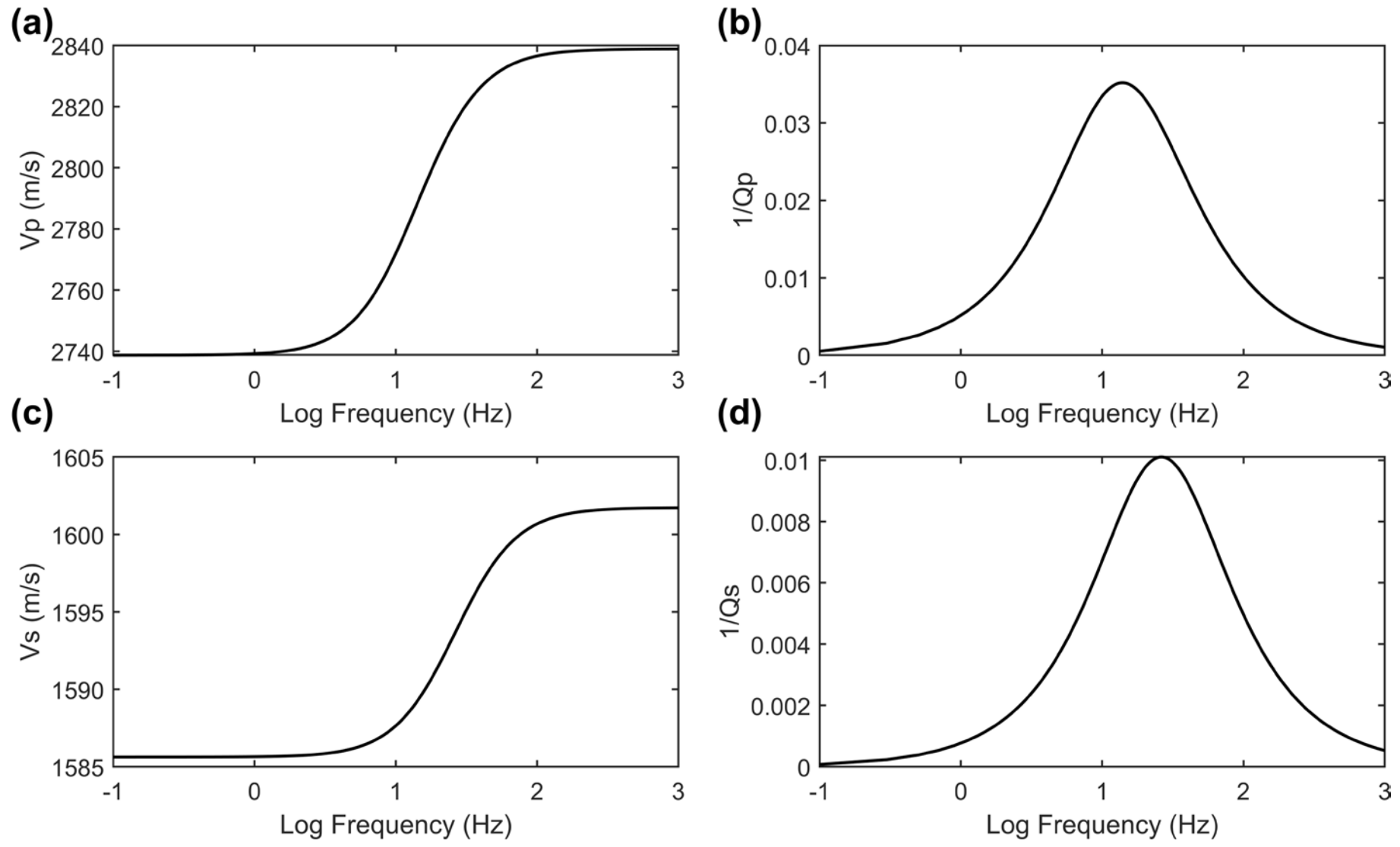

Fig. B1. P- and S-wave phase velocities and attenuations at the poroelastic interface for crack density of 0.05.

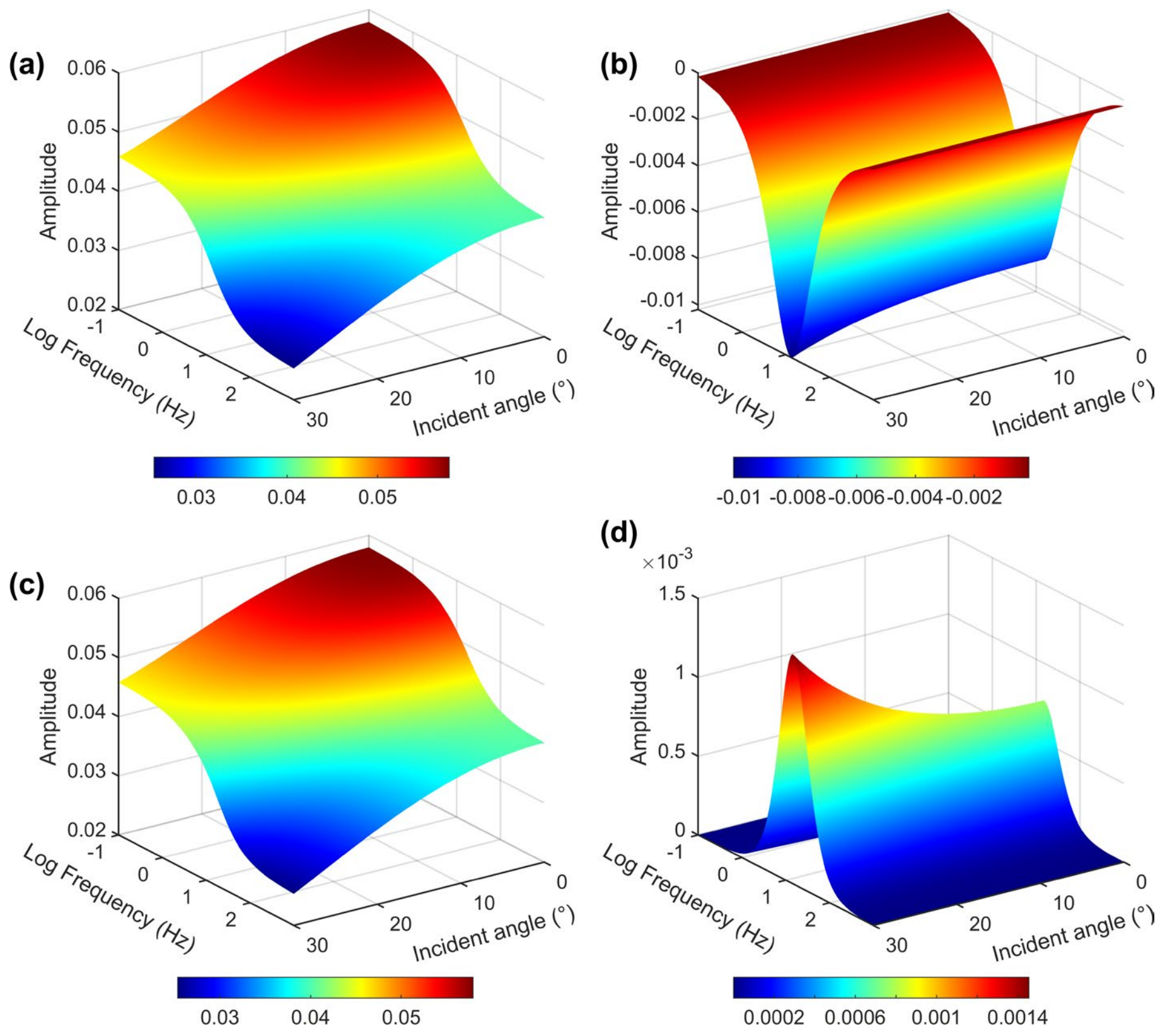

Fig. B2. (a) Real part, (b) imaginary part, (c) modulus, and (d) difference between the modulus and the real part of the complex reflection coefficient in Eq. 10 at crack density of 0.05.

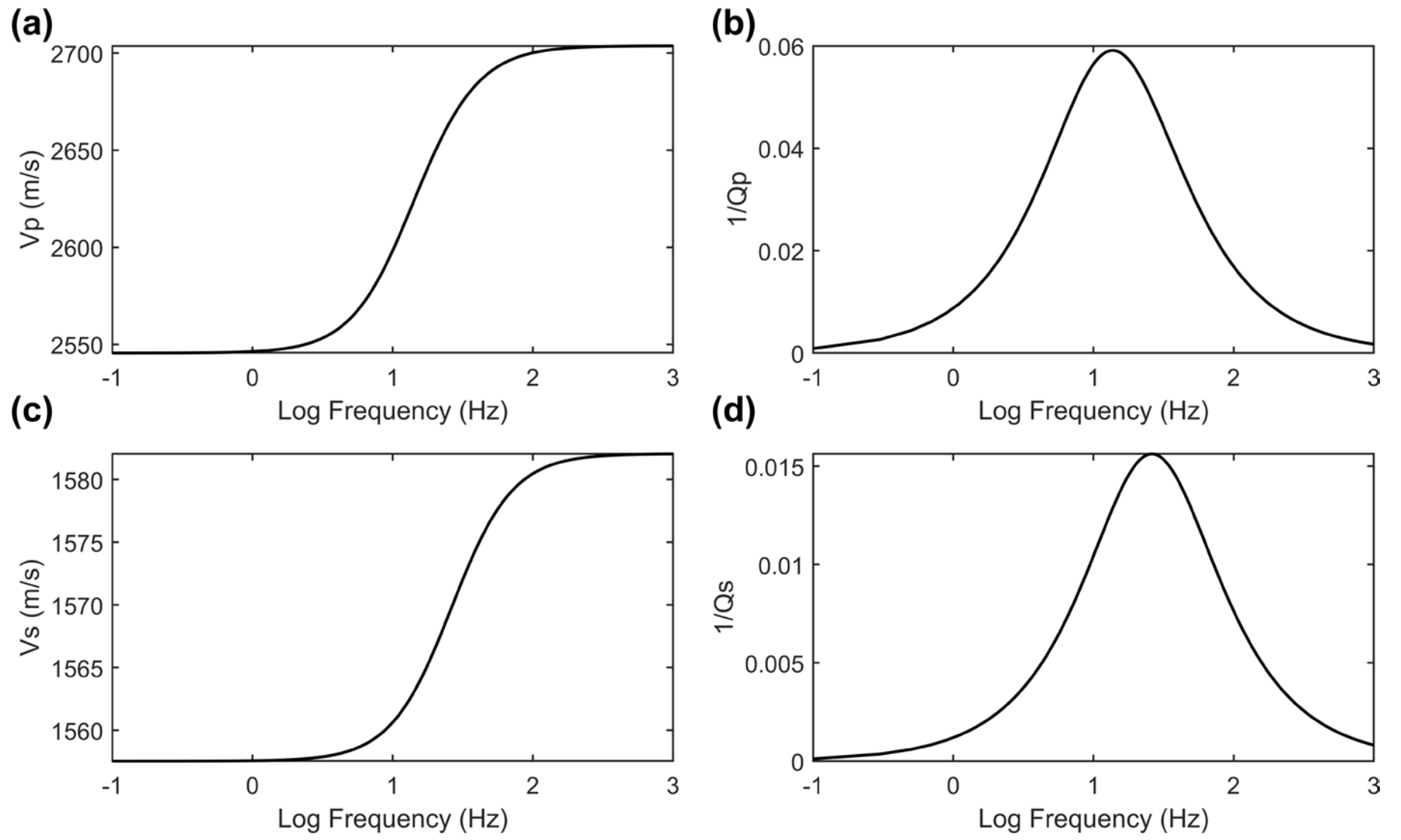

Fig. B3. P- and S-wave phase velocities and attenuations at the poroelastic interface for crack density of 0.05.

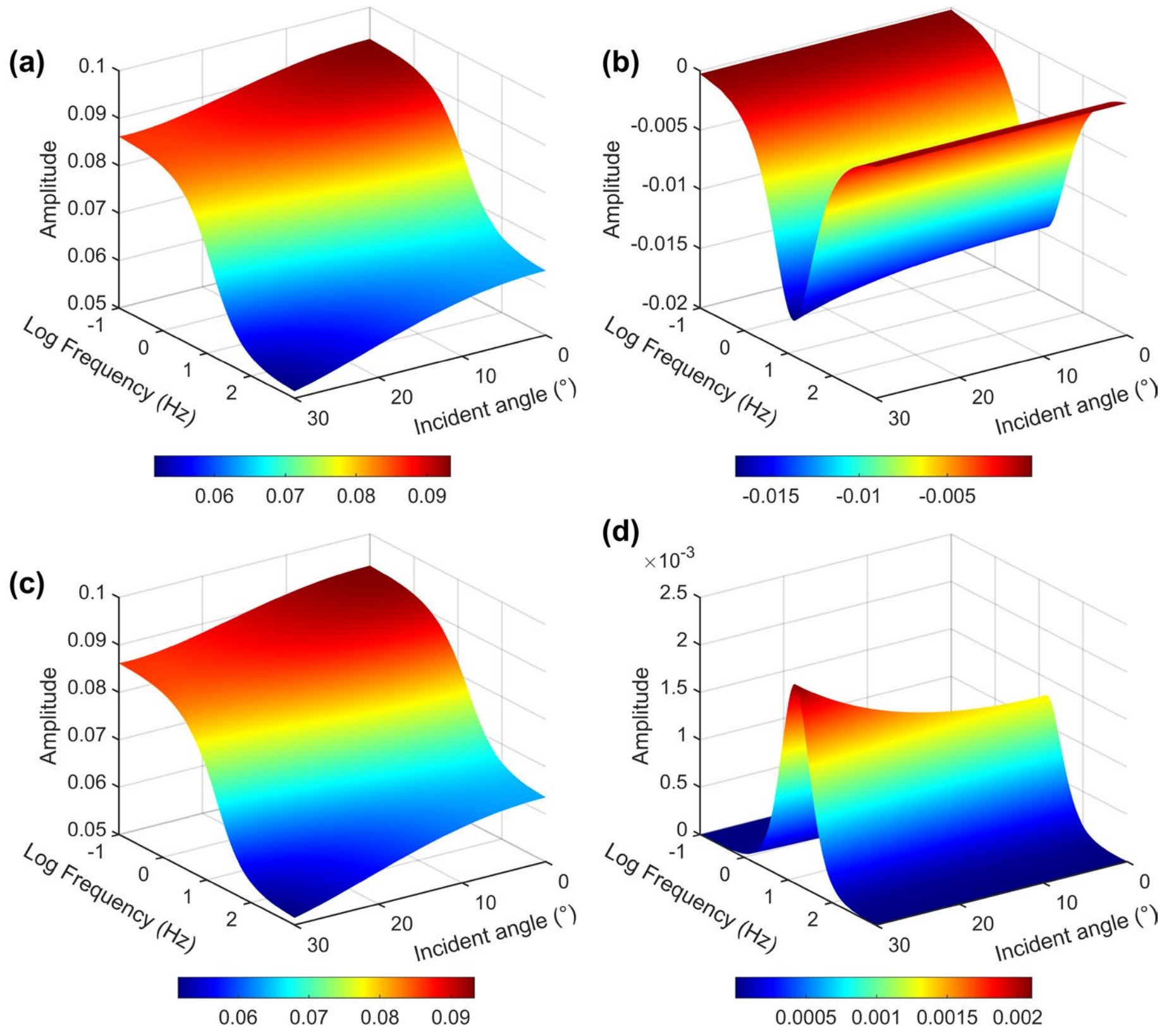

Fig. B4. (a) Real part, (b) imaginary part, (c) modulus, and (d) difference between the modulus and the real part of the complex reflection coefficient in Eq. 10 at crack density of 0.05.

## Appendix C: Estimation of P- and S-wave attenuations from well logs using Dvorkin–Mavko attenuation model

The siesmic attenuation can be calculated by:

$$Q^{-1} \approx \frac{M_H - M_L}{2\sqrt{M_H M_L}}, \tag{C1}$$

where $M_H$ and $M_L$ are the high frequency and low frequency limits of the seismic modulus.

For partial saturation attenuation caused by fluid heterogeneity, the rock is in a relaxed state in the low-frequency limit, and the pore fluid can be treated as a homogeneous water-gas mixture. In this case, the P-wave modulus equals that obtained from Gassmann's fluid substitution for the mixed fluid:

$$M_{L_partial} = K_{dry} + \frac{\left(1 - K_{dry}/K_m\right)}{\phi/K_f + \left(1-\phi\right)/K_m - K_{dry}/K_m^2} + \frac{4}{3}G_{dry}\,, \tag{C2a}$$

$$K_f = \frac{S_w}{K_w} + \frac{\left(1 - S_w\right)}{K_g}, \tag{C2b}$$

where $K_{dry}$ is dry-rock bulk modulus, $K_m$ and $G_{dry}$ are the bulk modulus of the rock matrix and shear modulus of the dry-rock, $\phi$ is porosity, $S_w$ is water Saturation, $K_w$ and $K_g$ are bulk modulus of the water and gas.

At the high-frequency limit, the equivalent P-wave modulus of the partially saturated rock can be regarded as the harmonic mean of the P-wave moduli of the water-saturated and gas-saturated rocks:

$$\frac{1}{M_{H_partial}} = \frac{S_w}{M_{water}} + \frac{1-S_w}{M_{gas}}, \tag{C3}$$

where $M_{water}$ and $M_{gas}$ are the P-wave moduli of the water-saturated and gas-saturated rocks, respectively, obtained from Gassmann's equations for water and gas fluid substitution.

Substituting Eqs. C3 and C2b into Eq. C1 yields the P-wave attenuation under partial saturation:

$$Q_{p_partial}^{-1} \approx \frac{M_{H_partial} - M_{L_partial}}{2\sqrt{M_{H_partial}M_{L_partial}}}\,. \tag{C4}$$

For background attenuation caused by elastic heterogeneity in saturated rocks, the P-wave modulus in the low-frequency limit is:

$$M_{L_back} = \bar{K}_{dry} + \frac{\left(1-\bar{K}_{dry}/\bar{K}_m\right)}{\bar{\phi}/K_w + \left(1-\bar{\phi}\right)/\bar{K}_m - \bar{K}_{dry}/\bar{K}_m^2} + \frac{4}{3}\bar{G}_{dry}, \tag{C5}$$

where $\bar{K}_{dry} = \left\langle K_{dry}^{-1} \right\rangle^{-1}$ and $\bar{G}_{dry} = \left\langle G_{dry}^{-1} \right\rangle^{-1}$ are the harmonic means of the dry-rock bulk and shear moduli, $\bar{K}_m = \left\langle K_m^{-1} \right\rangle^{-1}$ is the harmonic mean of bulk modulus of the rock matrix, and $\bar{\phi} = \langle \phi \rangle$ is the arithmetic mean of the porosity.

The P-wave modulus in the high-frequency limit is:

$$M_{H_back} = \left\langle \left[ K_{dry} + \frac{\left(1-K_{dry}/K_m\right)}{\phi/K_f + \left(1-\phi\right)/K_m - K_{dry}/K_m^2} + \frac{4}{3}G_{dry} \right]^{-1} \right\rangle^{-1}. \tag{C6}$$

Substituting Eqs. C5 and C6 into Eq. C1 yields the background P-wave attenuation:

$$Q_{p_back}^{-1} \approx \frac{M_{H_back} - M_{L_back}}{2\sqrt{M_{H_bcak} M_{L_bcak}}}. \tag{C7}$$

The total P-wave attenuation is:

$$Q_p^{-1} = Q_{p_back}^{-1} + Q_{p_partial}^{-1}. \tag{C8}$$

In the isotropic case, the shear wave attenuation is calculated as follows:

$$Q_s^{-1} = \left[ \frac{2\gamma}{\left(3\gamma-2\right)} + \frac{\gamma}{3\left(\gamma-1\right)} \right] \Big/ \left[ \frac{5}{4} \frac{\left(\gamma-2\right)^2}{\left(\gamma-1\right)} \right] Q_{p_back}^{-1}, \tag{C9}$$

where $\gamma$ is the squared ratio of the P-wave velocity to the S-wave velocity in the water-saturated rock.

# Appendix D: Explicit Form of the Matrix in Eq. 22

$$\mathbf{G}_k = \begin{bmatrix} g(\omega_k, t_1) & g(\omega_k, t_2) & \cdots & g(\omega_k, t_N) \\ g(\omega_k, t_2) & g(\omega_k, t_1) & \cdots & g(\omega_k, t_{N-1}) \\ \vdots & \vdots & \ddots & \vdots \\ g(\omega_k, t_N) & g(\omega_k, t_{N-1}) & \cdots & g(\omega_k, t_1) \end{bmatrix}, \tag{D1}$$

$$\mathbf{M}_k = \begin{bmatrix} e^{-i\omega_k t_1} & 0 & \cdots & 0 \\ 0 & e^{-i\omega_k t_2} & \cdots & 0 \\ \vdots & \vdots & \ddots & \vdots \\ 0 & 0 & \cdots & e^{-i\omega_k t_{N_t}} \end{bmatrix}, \tag{D2}$$

$$\mathbf{W}_j = \begin{bmatrix} W_j(\omega_1) & 0 & \cdots & 0 \\ 0 & W_j(\omega_2) & & 0 \\ \vdots & 0 & \ddots & \vdots \\ 0 & 0 & \cdots & W_j\left(\omega_{N_f}\right) \end{bmatrix}, \tag{D3}$$

$$\mathbf{P} = \begin{bmatrix} e^{-i\omega_1 t_1} & e^{-i\omega_1 t_2} & \cdots & e^{-i\omega_1 t_{N_t}} \\ e^{-i\omega_2 t_1} & e^{-i\omega_2 t_2} & \cdots & e^{-i\omega_2 t_{N_t}} \\ \vdots & \vdots & \ddots & \vdots \\ e^{-i\omega_{N_f} t_1} & e^{-i\omega_{N_f} t_2} & \cdots & e^{-i\omega_{N_f} t_{N_t}} \end{bmatrix} \tag{D4}$$

$$\mathbf{\Lambda}_j = \left[ A_j \mathbf{I}_{N_f}, B_j \mathbf{I}_{N_f}, C_j \mathbf{I}_{N_f}, \mathbf{D}_j, \mathbf{E}_j \right], \tag{D5}$$

$$\begin{aligned} \mathbf{D}_j &= diag\left( \frac{1}{2\pi \cos^2 \theta_j} \ln \frac{\omega_l}{\omega_0} \right)_{l=1,2\ldots N_f}, \\ \mathbf{E}_j &= diag\left( -\frac{4\sin^2 \theta_j}{\pi} \frac{Z_s^2(\omega_0)}{Z_p^2(\omega_0)} \ln \frac{\omega_l}{\omega_0} \right)_{l=1,2\ldots N_f}, \end{aligned} \tag{D6}$$

$$\begin{aligned} A_j &= \frac{1}{2}\left(1 + \tan^2 \theta_j\right), \\ B_j &= -4\sin^2 \theta_j \frac{Z_s^2(\omega_0)}{Z_p^2(\omega_0)}, \\ C_j &= -\left[ \frac{1}{2} \tan^2 \theta_j - 2\sin^2 \theta_j \frac{Z_s^2(\omega_0)}{Z_p^2(\omega_0)} \right], \end{aligned} \tag{D7}$$

$$\begin{aligned}
&\mathbf{m} = \left[ \mathbf{m}_{Z_p}; \mathbf{m}_{Z_s}; \mathbf{m}_{\rho}; \mathbf{m}_{Q_p^{-1}}; \mathbf{m}_{Q_s^{-1}} \right] \\
&\mathbf{m}_{Z_p} = \left[ \ln Z_p(t_1), \ln Z_p(t_2), \cdots \ln Z_p(t_{N_t+1}) \right]^T \\
&\mathbf{m}_{Z_s} = \left[ \ln Z_s(t_1), \ln Z_s(t_2), \cdots \ln Z_s(t_{N_t+1}) \right]^T \\
&\mathbf{m}_{\rho} = \left[ \ln \rho(t_1), \ln \rho(t_2), \cdots \ln \rho(t_{N_t+1}) \right]^T \\
&\mathbf{m}_{Q_p^{-1}} = \left[ Q_p^{-1}(t_1), Q_p^{-1}(t_2), \cdots Q_p^{-1}(t_{N_t+1}) \right]^T \\
&\mathbf{m}_{Q_s^{-1}} = \left[ Q_s^{-1}(t_1), Q_s^{-1}(t_2), \cdots Q_s^{-1}(t_{N_t+1}) \right]^T
\end{aligned}, \tag{D8}$$

$$\mathbf{I}_{N_f} = \begin{bmatrix} 1 & 0 & \cdots & 0 \\ 0 & 1 & \cdots & 0 \\ \vdots & \vdots & \ddots & \vdots \\ 0 & 0 & \cdots & 1 \end{bmatrix}_{N_f \times N_f}, \mathbf{I}_5 = \begin{bmatrix} 1 & 0 & 0 & 0 & 0 \\ 0 & 1 & 0 & 0 & 0 \\ 0 & 0 & 1 & 0 & 0 \\ 0 & 0 & 0 & 1 & 0 \\ 0 & 0 & 0 & 0 & 1 \end{bmatrix}, \mathbf{D}_0 = \begin{bmatrix} -1 & 1 & 0 & \cdots & 0 \\ 0 & -1 & 1 & \cdots & 0 \\ \vdots & \vdots & \ddots & \ddots & \vdots \\ 0 & 0 & \cdots & -1 & 1 \end{bmatrix}. \tag{D9}$$